\documentclass[11pt]{article}
\usepackage[utf8]{inputenc}
\usepackage[english]{babel}
\usepackage{a4wide}
\usepackage{authblk}
\usepackage{fancyhdr}
\usepackage{amsmath}
\usepackage{mathptmx}
\usepackage{booktabs}
\usepackage{rotating}
\usepackage{multirow}
\usepackage{tabularx}
\usepackage{makeidx}
\usepackage{graphicx}
\usepackage{multicol}
\usepackage[bottom]{footmisc}
\usepackage{upgreek}
\usepackage{natbib}
\usepackage{longtable}

\newcommand{\du}[0]{\mathrm{d}}

\newcommand{\aff}[0]{\uppsi}

\newcounter{sidenote}

\author[1]{Telmo Menezes\thanks{menezes@cmb.hu-berlin.de}}
\author[1]{Camille Roth\thanks{roth@cmb.hu-berlin.de}}
\affil[1]{Centre Marc Bloch Berlin (An-Institut der Humboldt Universit\"at, UMIFRE CNRS-MAE),
Computational Social Science Team, Friedrichstr. 191, 10117 Berlin, Germany}

\date{}

\begin{document}

\title{Automatic Discovery of Families\linebreak of Network Generative Processes}

\maketitle

\begin{abstract}
Designing plausible network models typically requires scholars to form a priori intuitions on the key drivers of network formation.  Oftentimes, these intuitions are supported by the statistical estimation of a selection of network evolution processes which will form the basis of the model to be developed.  Machine learning techniques have lately been introduced to assist the automatic discovery of generative models. These approaches may more broadly be described as "symbolic regression", where fundamental network dynamic functions, rather than just parameters, are evolved through genetic programming.
This chapter first aims at reviewing the principles, efforts and the emerging literature in this direction, which is very much aligned with the idea of creating artificial scientists.  Our contribution then aims more specifically at building upon an approach recently developed by us [Menezes \& Roth, 2014] in order to demonstrate the existence of families of networks that may be described by similar generative processes. In other words, symbolic regression may be used to group networks according to their inferred genotype (in terms of generative processes) rather than their observed phenotype (in terms of statistical/topological features). Our empirical case is based on an original data set of 238 anonymized ego-centered networks of Facebook friends, further yielding insights on the formation of sociability networks.
\end{abstract}

\section{Introduction}
\label{sec:introduction}

Networks have become over the last decades a key notion for modeling systems in a wide variety of fields. This is especially so in social sciences where networks are being introduced in an increasing number of contexts. On one hand, they are a type of abstraction that lends itself very naturally to the representation of a great variety of social structures and interactions. On the other hand, the information technology revolution has been making networks both more explicitly present -- for example due to the popularity of online social media -- and easy to retrieve by researchers.

Being practitioners in the field of \emph{Computational Social Sciences}, we have been concerning ourselves for some years with the challenge of deriving explanatory models from such complex empirical data. Networks are typically generated by phenomena that are non-linear in nature. The complex interactions between actors and the emergent environment they create -- represented by the network itself -- make it difficult to employ \emph{divide-and-conquer} approaches, where the problem can be divided into smaller parts that become tractable for human researchers to reason about. In other words, it is not easy to \emph{intuit} network formation principles which translate into simple yet successful generative models. Our belief that it makes sense to recruit computational intelligence to overcome this challenge led us to develop a method to automatically propose plausible and understandable network generators -- mathematical expressions that describe how new links are formed in the network, using only local variables (e.g., the current degrees of the pair of nodes in a candidate connection). This is akin to a multi-agent system, sufficiently abstract to lend itself to the description of a variety of phenomena. In the article where we proposed the full method for the first time \citep{menezes2014symbolic}, we showed that it could be used to discover plausible and simple generators not only for social, but also biological and man-made networks.

In the last years, machine learning has been gaining popularity as a scientific tool among many fields, partly because of the recent successes in \emph{deep learning}. We use a different approach, coming from the Artificial Intelligence branch usually known as \emph{evolutionary computation}. More specifically, we use a \emph{genetic programming} approach, given that we are evolving computer programs. There are two main reasons for this choice: the nature of the problem and the goal of understandability.

\begin{figure}
\centering
\includegraphics[width=.4\linewidth]{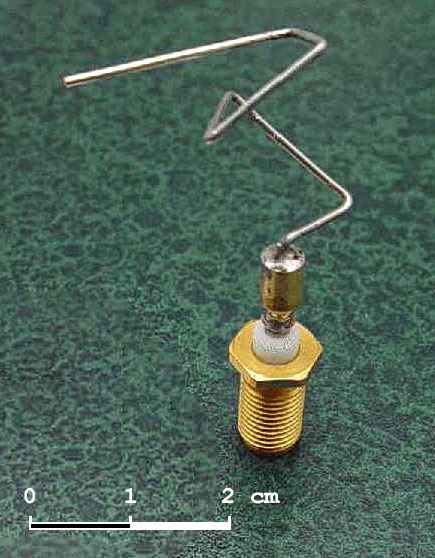}
\caption{This unconventional antenna design was generated by NASA using evolutionary computation to optimize its radiation pattern \citep{hornby-2006-automated}. It was used in the ST5 spacecraft. (Image in the public domain.)}
\label{fig:antenna}
\end{figure}

Many machine learning approaches, including the training of neural networks through back-propagation, require that an optimization criterion can be represented as a convex function, for which an optimum can be found through some form of gradient descent. The space of possible network generators appears too complex for such a convex function to be defined. In this kind of situation, evolutionary computation provides a stochastic and heuristic-driven approach to find viable solutions. The term ``evolutionary'' comes from its inspiration in Darwinian evolution. The simple principle of preference for the propagation of the most promising individuals with random mutations unleashes a type of intelligence that, although not human-like, is distinctly creative. To illustrate this, we show in figure~\ref{fig:antenna} an antenna created by NASA, that was designed by an evolutionary computation algorithm, aiming at optimizing its radiation pattern. We were interested in this ability to effectively explore a complex search space while being able to entertain \emph{counter-intuitive} solutions.

Another problem with many approaches such as neural networks if that they tend to be black
boxes. Even solving the convexity problem, they might produce good results in replicating network morphogenesis, yet they do not lend themselves to creating interpretable processes. We defined our genetic programs in a simple way, and included in our method a preference for simpler programs. As we will see, they can be translated into human-readable mathematical expressions. Our results are thus comparable with classical models of network morphogenesis, for which (human) scientists are however usually in charge of proposing plausible formation processes.

In this chapter we provide a wider view of our work, and also share new results. In the next section we discuss the last decades of research on the modeling of network morphogenesis, while providing a systematization that aims to help situate our work within it. We pay special attention to the recent history of evolutionary models, of which we were not the only pioneers.

In section~\ref{sec:symbolic} we provide a synthetic description of our method of symbolic regression of network generators. For all the details, we invite the reader to refer to our original article.

In section~\ref{sec:families} we present the results of novel research, aiming at finding families of generators within a dataset of networks of the same nature -- in this case, ego-centered friendship networks extracted from Facebook. We were interested in finding if symbolic regression would lead to sets of similar explanations. In other words, while network families are traditionally based on phenotypical resemblance \citep[see e.g.,][]{milo2004superfamilies,costa-2007-characterization,estrada-2007-topological,guimera-2007-classes,onnela-2012-taxonomies,avena-koenigsberger-2015-network}, we show here that our approach can yield families of generative processes, at the level of genotypal resemblance. We propose a new way to measure generator similarity, allowing us to project all the generators into a two-dimensional embedding, where generators with similar behaviors tend to be closer. With the help of this embedding, we were able to identify general patterns that many of the generator expressions conform to.
From a sociological perspective, we thus also shed light on a variety of plausible mechanisms of formation of ego-centered friendship networks.
More broadly, the existence of generator families further validates the behaviors embedded in the general mathematical expression characterizing a given family since it is able to efficiently reproduce the shape of several empirical networks.

\section{Network morphogenesis}\label{sec:gmorpho}

To illustrate the complexity of the task of intuiting efficient generative principles, we shall first review the existing efforts in this area.  We thereby intend to show better where our approach may fit in and benefit this state of the art. This will enable us to emphasize the interface position occupied by our work, which aims at inferring formation processes from the network while at the same time reconstructing it, using evolutionary modeling to avoid positing prior assumptions on the shape of these processes.

The modeling of network morphogenesis has generated a substantial literature over the last decades, especially after the early 2000s when most real-world networks were shown to exhibit peculiar connectivity and modular features. 
The corresponding state of the art may essentially be organized according to two key dichotomies: the first one relates to the \emph{target} of models, the second one to their \emph{foundations}.  More precisely, \hbox{(i)} models aim at reconstructing either network evolution processes or morphology; and to that end, \hbox{(ii)} they rely on assumptions, or input, related either to processes or to morphology. This yields the straightforward double dichotomy shown on Table~\ref{tab:ddich}, which includes a few canonical examples. Let us start by reviewing each category of that dichotomy.

\begin{table}[!b]
\begin{center}
\begin{tabular}{l>{\em}p{.3cm}|p{4.9cm}p{.6cm}p{4.9cm}}
\toprule
&\multicolumn{1}{c}{}&\multicolumn{3}{c}{\sc \textbf{reconstructing}}\\
&\multicolumn{1}{c}{}&\multicolumn{1}{c}{\sc processes}&&\multicolumn{1}{c}{\sc structure}\\\cmidrule{3-5}
~\newline
\multirow{10}{*}{\begin{sideways}\sc\textbf{using}\end{sideways}}&~\newline \begin{sideways}{\sc ~processes}\end{sideways}&
Preferential attachment estimation, \newline  Link prediction, Classifiers, \newline Scoring methods, ...&&
Preferential attachment-based generative models, Rewiring, Cost optimization, Social Simulation,  Agent-Based Models (ABMs), ...\\
&~\newline \begin{sideways}\sc ~structure\end{sideways}&
~\newline Exponential Random Graph Models (ERGMs), $p_1$, $p^*$,  Markov Graphs, Stochastic Actor-Oriented Models (SAOMs), 
...
&&
~\newline Prescribed structure,\newline Subgraph-based constraints, \newline Kronecker graphs, Edge swaps, ...\smallskip\\
\bottomrule
\end{tabular}
\caption{\label{tab:ddich}Double dichotomy of canonical network modeling approaches, which generally aim at reconstructing either evolution processes or network structure, and do so by relying either on evolution processes or network structure.}
\end{center}
\end{table}

\subsection{Reconstructing processes}

We first focus on the understanding of the generative processes at the lowest level, \hbox{i.e.} the rules governing the appearance or disappearance of nodes, and/or the formation or disruption of links. 

\subsubsection*{Using micro-level processes}
One of the most straightforward approaches to derive these rules consists in using, precisely, data describing these very dynamics, at the node and link level. In this category, we find simple counting methods aimed at appraising the propensity of links to form preferentially more towards nodes possessing certain properties --- this is the archetypal notion of ``preferential attachment'' (PA).  In its most restrictive yet most widespread acceptation, PA relates to the ubiquitous observation that links tend to attach to nodes proportionally to their degree. Following \cite{soll:gene}, this acceptation essentially stems from \cite{bara:evol} and \cite{jeong:measurin}. Several authors extended this notion beyond degrees to deal with a variety of both structural and non-structural features, including spatial distance \citep{yook-2002-modeling}, common acquaintances or topological distance \citep{koss:empi}, similarity \citep{filippo-menczer-2004-evolution,roth-gene,leskovec-2008-planetary-scale} or a combination thereof \citep{coin:loca}.
A more recent stream of research took this approach the other way around by proposing normative growth process and comparing them with empirical link formation. For one, \cite{papadopoulos-2012-popularity} introduced a model of link creation based on a concept of geometric optimization: nodes are placed in a plane and new nodes may connect to a subset of existing nodes by minimizing a geometric quantity. The model thereby reproduces connection probabilities observed in a selection of real networks, rather than observing real data to infer connection probabilities. 

Approaches inspired by machine learning have also been proposed to abstract processes by observing processes. They principally aim at predicting the appearance of links by generalizing from past link creation. This stream is rather geared towards prediction success rather than behavior estimation, \hbox{i.e.} efficiently guessing which links will appear rather than providing explicit link formation rules \cite[see][for a discussion of the relative performance of these methods]{yang-2015-evaluating}.  Scoring methods are among the simplest of these approaches: \cite{libe-link} first introduced a predictor function based on some dyadic feature (such as the number of common neighbors, Jaccard coefficients, Katz' distance). This function produces a ranking on non-connected dyads from the observation of an empirical network formed over the learning period $[t_0,t]$. The prediction task then consists in going through the dyad list in descending order and comparing it with the links that empirically appeared during a test period $[t,t']$. 

A large array of more sophisticated techniques have been used in this field, by involving, inter alia, SVM classifiers \citep[e.g., as proposed by][]{adar:impl} or more broadly supervised learning methods \citep{hasan-2006-link}, as well as matrix and tensor factorization \citep{acar-2009-link} (see \cite{lu-2011-link} and \cite{Hasan2011} for introductory reviews of this type of endeavors). 
Some authors divide the network into modules, or blocks, in order to estimate a simple (and local) probability of link formation within and between these modules, e.g., \cite{guimera-2009-missing} who define modules through stochastic blockmodeling \citep{ande:buil}, or \cite{clauset-2008-hierarchical} who use a dendrogram to both build the module partition and compute the inter-module connection probabilities.
Overall, there has been an increasing attention to the time-related and spatial variability of the prediction task by considering the local neighborhood of nodes, both in a topological and temporal manner \citep{sarkar-2014-nonparametric} and in a semantic fashion (e.g., by enriching the set of prediction features with content \citep{rowe-2012-who-will} or so-called sentiment analysis \citep{yuan-2014-exploiting}).
Also of note is the recent addition of evolutionary algorithms to this toolbox: for instance, \cite{bliss-2014-an-evolutionary} evolve a weight matrix describing the relative contributions of various similarity measures in predicting new connections.

\subsubsection*{Using macro-level structure}
Link formation principles may also be infered from the observed network topology. The most common approach in this stream comes to econometric techniques aimed at fitting a model whose parameters are associated with specific link formation effects and which takes the whole network as an input.

Exponential Random Graph Models (ERGMs) famously belong to this class. In all generality, they rely on the assumption that the observed network has been randomly drawn from a distribution of graphs. The probability of appearance of a given graph is construed as a parameterization on a choice of typical network formation processes: be they structural (such as transitivity, reciprocity, balance, etc.) or non-structural (such as gender dissimilarity, homophily, etc.). The aim is generally to find parameters maximizing the likelihood of the observed network. Each parameter then describes the likely contribution of the corresponding category of link formation process (\hbox{e.g.}, strong transitivity, weak reciprocity).
ERGMs have been introduced by \cite{holl:expo} through the so-called $p_1$ model describing the probability of graph $G$ as $p_1(G) \sim \exp(\sum_i \lambda_i v_i(G) ) = \Pi_i \exp(\lambda_i v_i(G))$ where $v_i(G)$ denotes a value related to the i-th process (\hbox{e.g.}, transitivity) and the $\lambda_i$ are the above-evoked parameters. $p_1$ assumes independence between dyads, which limits the model to simple dyad-centric observables: principally, degree and reciprocity. It can nonetheless be applied to a partition of the network into subgroups \citep{fienberg-1985-statistical} or stochastic blockmodels \citep{holland-1983-stochastic,ande:buil}, which posits a block structure, \hbox{i.e.} the fact that distinct groups of actors, or ``blocks'', exhibit distinct connection behaviors; parameters are thus a function of blocks. 
\cite{frank-1986-markov} later introduced ``Markov graphs'', which takes into account dependences between edges and thus triads and simple star structures, and which was subsequently extended as the $p^*$ model \citep{wasserman-1996-logit,ande:prim,robi-intr}. 
Further generalizations to more complex graph structures have lately been proposed e.g., for so-called ``multi-level networks'' \citep{wang-2013-exponential,brennecke-2016-the-interplay}, which are essentially graphs with two types of nodes and three possible types of links (two intra-type and one inter-type).

When longitudinal data is available, network evolution may be construed as a stochastic process. \cite{holland-1977-a-dynamic} then \cite{wasserman-1980-analyzing} proposed to appraise network dynamics as a (continuous-time) Markov chain. They assumed that the probability of link appearance or disappearance depends on a limited set of (static) parameters representing the contribution of various structural effects, such as, again, reciprocity, degree. Networks observed at different points in time are used to fit these parameters. Albeit not directly affiliated with this framework, the approach of \cite{powe-grow} proceeds in a similar fashion to determine the key factors guiding attachment of firms in a biotech sector. 
Stochastic actor-oriented models (SAOMs) further extend these ideas by introducing an actor-level viewpoint whereby actors establish link to optimize some objective function \citep{snij-stat}. Again, the parameters of this function denote effects deemed important for link formation (or destruction). These models also accommodate for some form of dyadic dependence, and take into account non-structural features (including gender). They may include behavioral observables \citep{snij-mode} or rely further on machine learning techniques e.g., by extending SAOMs to a Bayesian inference scheme \citep{koskinen-2007-bayesian}. 
In practice, SAOMs may be used to study non-structural effects linked to gender, racial, socioeconomic or geographical homophily, as demonstrated for instance in an online context on Facebook friendship \citep{lewi-soci}.
ERGMs and SAOMs assuredly share several traits, and it is also possible to develop ERGMs in a longitudinal framework as temporal ERGMs (or TERGMs), where the estimation for a graph at time $t$ depends on the graph at $t-1$ \citep{hanneke-2010-discrete}. 
For a more detailed comparison between SAOMs and ERGMs, see \cite{block-2016-forms,per-block-2018-change}.

On the whole, the advantage of these approaches over the previous process-based methods lies in the joint and concurrent appraisal of a variety of effects (each statistical model may consider an arbitrary number of variables to explain the shape of the observed network), with the drawback of reducing the contribution of each effect to a scalar quantity.


\subsection{Reconstructing structure}

The second part of the double dichotomy  (right-side on Table~\ref{tab:ddich}) relates to understanding the morphogenesis of the network itself. It may again be roughly divided into two broad categories, depending on whether approaches are based on 
a given growth process or on the topology of the network itself.

\subsubsection*{Using processes}

A myriad of models have been proposed to reconstruct network structure from normative assumptions. This is perhaps the most well-known and natural approach in statistical physics. At the core of these approaches lies generally a master equation or a master process featuring a certain number of key and oftentimes stylized ingredients. These ingredients correspond to an ideally small subset of canonical growth processes, defining the essential rules for adding --and, rarely, removing-- nodes, links, and most importantly towards which types of nodes.
The goal often consists in reproducing the observed connectivity (such as degree distributions), cohesiveness (such as clustering coefficients) or connectedness (such as component size distributions).

One of the earliest successful attempts at summarizing network morphogenesis with utterly simple processes consisted again in analytically solving simple PA based on node degree \citep{bara-emer}. Models based on a general notion of PA have been extended in various directions: taking into account the age of nodes \citep{doro-agin}, their Euclidean distance \citep{yook-2002-modeling,guimera-2004-modeling}, their intrinsic fitness \citep{cald:scal}, their rank \citep{fortunato-2006-scale-free} or their activity \citep{perra-2012-activity}; formalizing a notion of competition between nodes to attract new links \citep{fabr-heur,berg:comp,dsouza-2007-emergence}; copying links from ``prototype'' nodes \citep{kuma:stoc} or using random walks \citep{vazq:grow}; introducing preferences for transitive closure \citep{holme-2002-growing} or for specific groups of nodes (based on an a priori taxonomy \citep{lesk:grap} or an affiliation network \citep{zheleva-2009-co-evolution}); or mixing structural PA with semantic PA (e.g., \cite{filippo-menczer-2004-evolution} who introduces the so-called ``degree-similarity'' model after observing that connected web pages are rather more similar, or \cite{roth-phd} who mixes group-based PA and semantic PA).
Group-based PA may also be found in models which describe the addition of groups rather than dyadic links, such as \cite{guim:team}: the network evolves through the iterative addition of teams and thus links between all their members, assuming a certain propensity to introduce newcomers and repeat past interactions.

Another class of models is based on link rewiring. One of the simplest versions was introduced by \cite{watt-coll}, who start with a ring lattice of fixed degree and reconnect links with a given probability $p$. This led to a discussion of the resulting structure in terms of low path length and high clustering coefficient, or ``small-world''.  \cite{coli-netw} later reproduced these two statistical features by adopting a distinct approach based on a rewiring process aimed at optimizing a global cost function, in a way inspired by \cite{fabr-heur}.

Finally, a broad class of network models, especially in the social realm, falls into the category of \emph{agent-based models} as soon as they rely on a relatively rich combination of processes. They generally aim at a specific application field which, in turn, requires detailed assumptions: as such, they typically offer a good combination of realism (they benefit from a stronger sociological grounding) and tractability (their study generally requires to resort to simulation).
Examples of sophisticated models have been abundant in the social simulation 
literature from early on and are now present in a wide array of works at the interface with statistical physics and computational social science.  It is way beyond the scope of this paper to attempt an overview of the wide diversity of agent-based network models. Let us nonetheless casually mention \cite{gilbert-simulation}, who models the heterogeneous distribution of papers authored by scientists in a given field, and further reproduces the clustering of nodes in a semantic space, based on simple copying rules and the notion of quanta of knowledge called `kenes', by analogy with genes; \cite{pujol-2005-how-can-social}, who build various social exchange network shapes by combining various agent decision heuristics and cognitive constraints; and \cite{goet-mode} who reproduce blogger posting behavior and citation networks through a combination of random-walk-based generators and post selection rules.

\subsubsection*{Using structure}

Reproducing graph structure directly from graph structure essentially means showing that some structural constraints entail the presence of other structural features --- for instance by demonstrating that a certain number of connected components or a strong proportion of some sort of triads follows from a given degree or subgraph distribution. Early attempts precisely focused on prescribing a power-law degree sequence \citep{aiello-2000-a-random} and, shortly thereafter, any degree sequence \citep{newm:rand}. Several methods have later been proposed in the case of more sophisticated constraints, such as prescribed degree correlations \citep{maha-syst}, subgraph distributions \citep{karr-rand}, or recursive stuctures \citep{lesk-kron}.  

A typical challenge consists in being able to sample the space of graphs induced by a given set of constraints.
Some approaches manage to provide a closed-form expression of several average statistical properties of the induced graph space, as has been done for the typical path length or average clustering coefficient by \cite{newm:randsoci}. When this is not possible, an alternative consists in sampling the graph space through iterative exploration: the initial empirical graph is typically transformed by swapping pairs of edges while respecting the original constraint \citep{rao-1996-a-markov,gkan:mark}. This corresponds to a navigation in a meta-graph gathering all graphs of the target space. Beyond simple constraints, exhaustive navigation is usually impossible. \cite{tabo-gene
} practically address this issue with an empirical sampling method denoted as ``k-edge switching'', iteratively swapping groups of ($k$) links in order to cover an increasingly large portion of a given graph space.

\subsection{Combining both: evolutionary models}

In all four positions of the double dichotomy, the challenge generally consists in proposing one or several processes or constraints which will be key to explain network formation --- be it transitivity, centrality, homophily, etc. 
The importance of such and such mechanism may be either assumed \emph{a priori}, by looking at its effect on the network evolution, or verified \emph{a posteriori}, by confirming its existence and appraising its shape during the network evolution. 
In all cases, intuition plays a key role. Yet, creating these models requires insights that may sometimes be unconventional.

To alleviate this dependence, evolutionary algorithms were recently used to automatically propose sets of mechanisms inferred from the observed structure. It differs from the above-mentioned methods in that it jointly uses the structure to reconstruct processes and the processes to reconstruct the structure. More precisely, network structure is used to devise link formation processes and, in turn and iteratively, these discovered processes are precisely used to reconstruct the structure.

Some of the earlier approaches introduced template models based on sets of possible specific actions (e.g., creating a link, rewire an edge, connecting to a random node, etc.). Actions have been organized in various manners: first as a fixed chart, resembling the typical structure of agent-based models \citep{menezes-2011-evolutionary}, as a sequential list of variable size \citep{bailey-2012-automatic,bail-gene,harrison-2015-investigating,harrison-2016-a-meta-analysis} or, very recently, as a matrix whose weights describe the relative contribution of each action \citep{arora-2016-a-multi-objective,arora-2017-action-based}. In all these works, the evolutionary process aims at automatically 1. filling the template model with actions and 2. fitting the corresponding parameters. As is typical in evolutionary programming, it involves a fitness function which evaluates the resemblance between the empirical network and networks produced by the evolved model. Fitness functions rely on classical structural features (degree distributions, motifs, distance profiles, etc.). Models are iteratively evolved along increasing fitness values. 

In parallel, we further proposed an original approach based on genetic programming and aimed at inferring arbitrarily complex combinations of elementary processes, construed as laws \citep{mene-auto,menezes2014symbolic}.\footnote{
In terms of potential applications, this approach has been evoked in the context of human connectome modeling~\citep{betzel2016generative, adolphs2015unsolved}, as an alternative to conventional social simulation models~\citep{amblard2015models} or to appraise matrimonial preferences from genealogical networks ~\citep{menezes2016new}.} We first introduced a generic vocabulary making it possible to describe network evolution in a unified framework, as an iterative process based on the likelihood of appearance of a link between two nodes, construed as a function on node properties in the currently evolving network (\hbox{i.e.} a form of generalized preferential attachment) --- relying on structural features such as distance, connectivity, as well as non-structural characteristics. Representing these functions as trees enabled us to apply genetic programming techniques to evolve rules which are then used to generate network morphologies increasingly similar to the target, empirical network. 

This technique may be denoted as ``symbolic regression'', for the goal is to use genetic programming to evolve free-form symbolic expressions rather than fitting parameters associated to fixed symbolic expressions: we \emph{automatically} evolve realistic morphogenetic rules from a given instance of an empirical network, thereby symbolically regressing it. This strategy is inspired by the work of \cite{schmidt-2009-distilling} who extract free-form scientific laws from experimental data.  We first applied our method on kinship networks \citep{mene-auto} which led to the publication of a much more general manuscript \citep{menezes2014symbolic}.
One remarkable result consists of the ability to \emph{systematically} and \emph{exactly} discover the laws of an Erd\H{o}s-R\'enyi or Barab\'asi-Albert generative process from a given stochastic instance. Distinct, realistic and compact laws for a variety of social, physical and biological networks could also be found. 

\smallskip
We now describe in detail the core of the symbolic regression approach.


\section{Symbolic Regression of Network Generators}
\label{sec:symbolic}
We construe network generation as a stochastic process where edges are added iteratively, following some probability-based preference. Our approach is embedded in a generalized preferential attachment framework centered around the notion of \emph{generator} which is a scoring function providing a way to prefer some link over the others.  A generator thus assigns a score $s_{ij}$ to all edges $(i,j)$. At each step of the network construction, a random sample $S$ of candidate edges is drawn, among which a new edge is stochastically selected with a probability $P_{ij}$ proportional to $s_{ij}$ such that: 
\begin{equation}
P_{ij} = \frac{s_{ij}}{\displaystyle\sum_{i',j' \in S} s_{i'j'}}
\end{equation}
In practice, we forbid negative values and replace them with 0; in the special case where all weights are zero, they are all set to $1$ for mathematical consistency.

In other words, generators implement an (arbitrarily complex) form of PA restricted to a random subset of links.  Our core aim thus consists in designing a process able to automatically discover score computation functions $s$ which yield networks comparable to a target empirical network.
We construe generators as \emph{tree-based computer programs} which represent mathematical expressions. 
Tree nodes are operators while leaves are variables and constants. Operators include classical arithmetic operations $\{+$, $-$, $*$, $/\}$, general-purpose mathematical functions: $\{x^y$, $e^x$, $\log$, $\text{abs}$, $\min$, $\max\}$, conditional expressions: $\{>, <, =, =0\}$ and an affinity function ($\aff$, which we will further describe below). Variables are classical monadic or dyadic network measures which apply to the two nodes participating in the edge $(i,j)$ to be scored: centrality degrees of each vertex ($k_i$ and $k_j$), topological distance between the two vertices ($\du$),\footnote{To compute distances we use on heuristic based on a random walk, for (1) the exact computation is computationally intensive and, what is more, (2) new connections are also likely guided by a hop-by-hop navigation mechanism instead of an omniscient knowledge of the exact number of hops separating two given nodes.} and their sequential identifiers ($i$ and $j$, whose role we also discuss later on).  We limit here the presentation of our approach to undirected networks with a fixed set of nodes, which fits our empirical material of Facebook ego-centered friendship networks. Nonetheless, it can straightforwardly be extended to directed networks (as in our original work) and the regular arrival of nodes.  

\begin{figure}[!th]
\includegraphics[width=\linewidth]{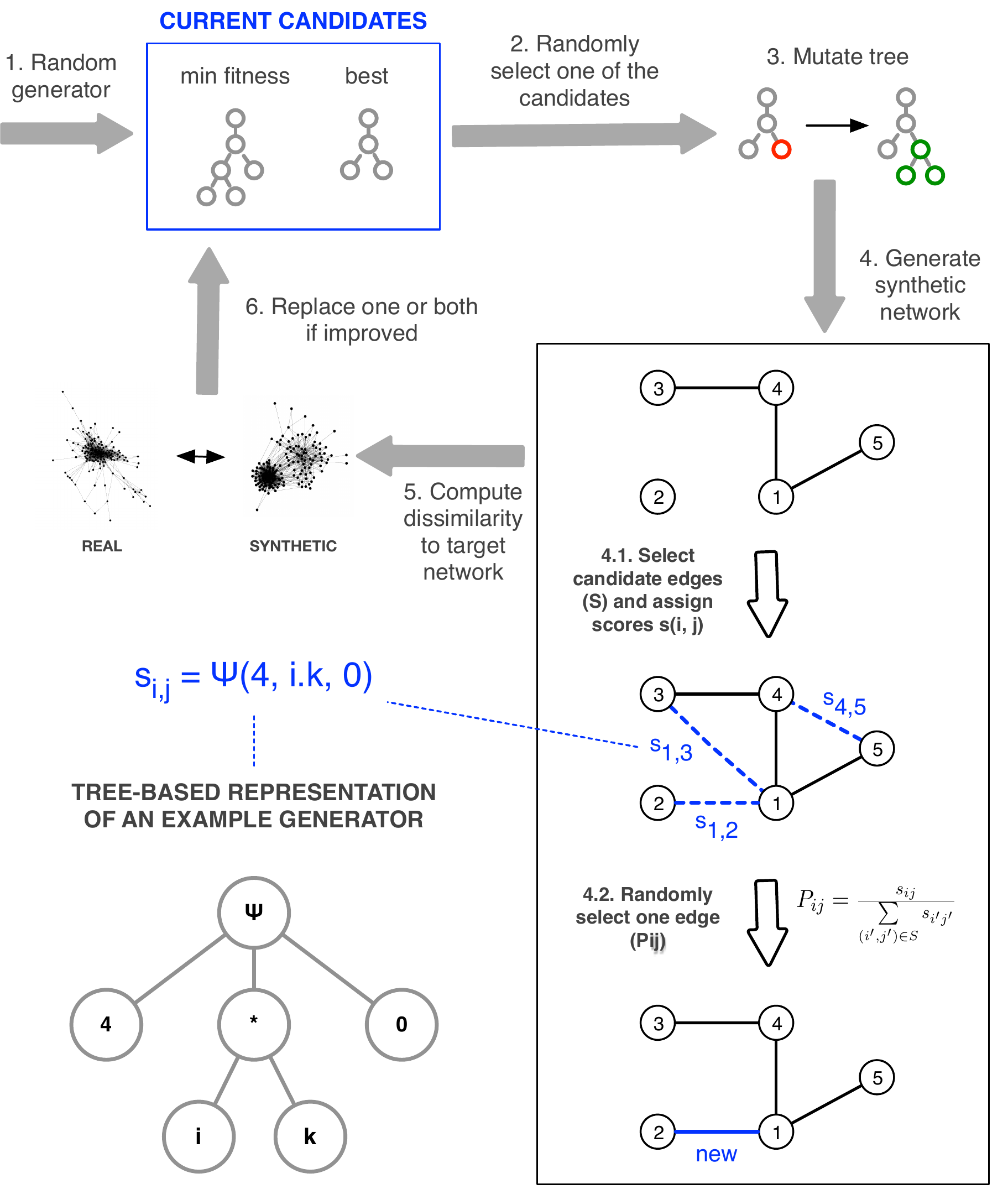}
\caption{Evolutionary loop including the synthetic network generation process. The outer part of this figure describes evolution at the generator population level, while the framed part on the right describes the evolution of a network for a given generator.}
\label{fig:generator_search}
\end{figure}

This simple setting provides a language for describing generators and expressions which model and produce non-linear and non-centralized growth mechanisms.  We now need a way to measure the similarity between the target network and generator-produced networks. This will provide the basis for defining the fitness function of our genetic approach.  To this end, we first use a combination of distributions related to various topological aspects of the network, such as degree and PageRank ~\citep{brin1998anatomy} centralities, distance distributions and triadic profiles~\citep{milo2004superfamilies}.  We then compute dissimilarities between the respective distributions: for centralities, we apply the Earth mover's distance (EMD)~\citep{ling2007efficient}, for the other distributions, we simply compute use a ratio-based dissimilarity metrics.  Of course, other metrics and dissimilarity measures may be used; we made these choices as a simple and intuitive trade-off between tractability and topological realism, which happens to work well.

By minimizing these dissimilarity measures, we get closer to the target network.  This corresponds to a multi-objective optimization problem where some dissimilarities have to be minimized to the possible expense of others. We adopt a simple strategy by considering all dissimilarities in regard to the improvement over a random network. In other words, we divide the dissimilarity between the target network and a generated network by the dissimilarity between the target network and the average of $30$ Erd\H{o}s-R\'enyi (ER) random networks of the same size (same number of nodes and edges as the target).
For a given metric, this means that if the dissimilarity between the target network and the ER average is, say, 5 and the distance from the target network to the generated network is 3, the ratio is 3/5. The smaller the ratio, the better the improvement --- a ratio of 1 corresponds to no improvement. 
The evolutionary algorithm then aims at improving generators by minimizing the highest of these ratios. This defines our fitness function: the lower its value, the better the fitness.\footnote{ER is admittedly a basic null model. Yet, opting for a richer model is likely to induce bias: for instance, a fitness function based on a comparison with a configuration model would precisely incorporate  target network degree distributions, making it impossible to directly approximate them.}


Our framework relies on a further feature: we allow node heterogeneity, \hbox{i.e.} we assume that not all nodes are and thus behave the same, irrespective of their structural position. Some actors of a social network may for example be intrinsically more likely to form ties with a specific class of actors. Here, we simply take heterogeneity into account through the sequential index of the node $i \in \{1,..,n\}$.
These indices, considered as \emph{identifiers}, may be used as a variable by a generator, and may thus introduce \emph{a priori} distinctions in actor types. As we shall see, this element is key in the case of friendship networks where social circles play an essential structuring role.

Consider for instance the generator $s(i, j) = \frac{1}{i}$. It induces a probability of edge creation entirely determined by the identifier of one of its extremities. Nodes have distinct \emph{a priori} propensities to originate connections, distributed following a hyperbolic curve. While integer identifiers may appear to introduce heterogeneity in very simplistic way, they can be combined with the other building blocks in an infinity of manners --- and our results below show that indices were indeed used in sometimes creative ways.

Furthermore, index-based heterogeneity may be leveraged to define generators where certain vertices have natural affinity for each other. This brings us to the \emph{affinity} function $\psi$, which uses the modulo operation to partition the identifier space into a certain number of groups. It relies on three operands: a constant, $g$, the number of groups; and two expressions, $a$ and $b$, which are conditional outputs. If target and origin nodes $i$ and $j$ are equal modulo $g$, and thus belong to the same group (\hbox{i.e.} in case of ``affinity''), the function returns $a$, and $b$ otherwise:

\begin{equation}
    \psi_g(i, j, a, b)= 
\begin{cases}
    a,    & \text{if } (i \bmod g) \equiv (j \bmod g)\\
    b,              & \text{otherwise},
\end{cases}
\end{equation}
 
From now on, we consider $i$ and $j$ as implicit variables and denote the function as: $\psi_g(a, b)$.

Combining all these elements into an evolutionary loop makes it possible to generate plausible models for network generators, as summarized on Fig.~\ref{fig:generator_search}. Several runs with the same target network may generate different models --- although they appear experimentally to converge onto similar behaviors. This leaves it to practitioners to select among the various options, conceivably by involving domain knowledge. A more objective consideration pertains to a trade-off between simplicity and precision. Since generators are essentially programs, complexity may be simply appraised through {program length}, an upper bound on the {Kolmogorov complexity}~\citep{ming1997introduction}. We thus apply a quantified version of {Occam's Razor}: all other things being equal, we choose the model with the lowest program length.

\section{Families of Network Generators}
\label{sec:families}


This approach provides the equivalent of an artificial scientist proposing plausible network models, replacing the intuition of the modeler. Using a biological analogy, it also makes it possible to discuss networks in terms of their plausible genotype (i.e., generator equations) rather than phenotypes (i.e., a series of topological traits).  

Phenotypical traits assuredly provide the basis for appraising the quality of structural reconstruction and, by extension, for defining fitness functions attentive to such and such topological property \citep[for an early yet already comprehensive review on the possible properties, see][]{costa-2007-characterization}. 
They also provide a good foundation for comparing networks with one another: a series of studies has indeed been devoted to defining network families by relying on triadic profiles \citep{milo2004superfamilies}, canonical analysis of various measures \citep[][section 19]{costa-2007-characterization}, adjacency matrix spectrum \citep{estrada-2007-topological}, blockmodeling \citep{guimera-2007-classes}, community structure \citep{onnela-2012-taxonomies}, hierarchical structure \citep{corominas-murtra-2013-on-the-origins}, communication efficiency \citep{goni-2013-exploring}, graphlets \citep{yaveroglu-2014-revealing}. Note that this last method has been precisely used by \cite{charbey-2018-graphlet-based} to phenotypically categorize the empirically networks we are dealing with here. Phenotypical traits have also been the target of evolutionary algorithms in \cite{martens-2017-symbolic}, who symbolically regress formulas describing the phenotype of the network, \hbox{e.g.} finding an explicit expression for the diameter of various classes of networks as a function of the number of nodes, links, or some eigenvalues of the adjacency matrix.

\bigskip
By contrast, symbolic regression enables the comparison and categorization of networks based on their plausible underlying morphogenesis rules --- as such a \emph{genotypic categorization}.  The core of the present contribution consists in applying our approach on a collection of networks of the same nature, unlike \cite{menezes2014symbolic} which addresses a limited number of networks of different natures --- biological, social and man-made, both directed and undirected. 

Here, we will exhibit families of generators, both in terms of their function and in terms of their expression. Their existence further suggests that a single mathematical expression and thus explanation may apply to a number of distinct empirical networks. In turn, it is thus even likely to correspond to a widespread class of actual generative behaviors.

\subsection{Protocol}

We use 238 anonymized ego-centered networks of Facebook friends which were randomly sampled from about 10,000 such networks collected in a large-scale online survey organized within a collaborative project called ``Algopol'' (consenting participants accepted to give access to their publication and network constitution history).
Unlike other social networks such as Twitter, with concepts of ``following'' and ``being followed'', Facebook friend relationships are reciprocal and thus undirected. Furthermore, in ego-networks, ego is by definition connected to every other node, so its presence would likely lead to more complex generators without any added explanatory power. We thus discard ego and all of their links.


For each network, we performed five evolutionary search runs. We then selected the generator discovered by the run that attained the highest fitness. This is a simple strategy to avoid low-quality local optima.

\subsection{A Measure of Generator Dissimilarity}

To identify families of generators and to visualize how similar they are in relation to each other, we start by introducing a measure of dissimilarity between pairs of generators. We understand the generator expression as the genotype and the network created using the generator as the phenotype. As in biology, different phenotypes can correspond to the same genotype. In our case, and beyond the intrinsic stochasticity of the generative process, this is trivially true because we can use the same generator to create networks of different sizes -- both in numbers of nodes and edges. It is also true that different genotypes can create similar phenotypes. Notions of dissimilarity could be imagined both on the genotype and phenotype sides. On the genotype side, this could be a measure of program dissimilarity, for example something akin to an edit distance. On the phenotype side, it could be a comparison of generated networks. We opted for the latter: we look for collections of generators that produce similar networks, and then check these groups to see if they contain regularities or competing explanations. In the end we propose a qualitative-quantitative analysis of families of generators.

Comparing networks is not a trivial task, and it becomes even harder for networks that do not have the same number of nodes and edges. With our generators, we are in a position to control this latter aspect. We use the generators to create synthetic networks that do not have the varied topologies of the ones they were derived from, but instead have a predetermined number of nodes and edges, facilitating subsequent comparison. We chose to generate networks of $1000$ nodes and $10000$ edges, deemed to be large and dense enough for comparisons to be meaningful, and yet not so large that the task of comparing all pairs would become computationally intractable.

For the comparison itself, we employ a modified version of the fitness function that was used during the generator discovery process. The fitness function for undirected networks uses four distribution distance measures: $k_d$ for degree; $PR_d$ for PageRank; $d_d$ for distance and $\tau_d$ for the triadic profile. In that case, these measures are used to compare a synthetic network against the target network. Here, we will use them to compare pairs of synthetic networks created by the discovered generators. Being $n=238$ the number of generators we consider four $n \times n$ matrices of pairwise distances, one for each measure: $D_k$, $D_{PR}$, $D_d$ and $D_\tau$. To make these measures directly comparable, we produce normalized versions of each of these metrics in the following way:

\begin{equation}
    D'_{i,j} = \frac{D_{i,j} - \min_{i'}(D_{i',j})}{\max_{i'}(D_{i',j}) - \min_{i'}(D_{i',j})} 
\end{equation}

The global dissimilarity function $\delta(i,j)$ is then simply the sum of the four normalized distances between two generated networks:

\begin{equation}
    \delta(i,j) = D'_{k_{i,j}} + D'_{{PR}_{i,j}} + D'_{d_{i,j}} + D'_{\tau_{i,j}}
\end{equation}

Notice that the above normalization process can lead to different estimations of the several distances depending on the direction, because the normalization process is not symmetrical. We therefore finally use a symmetrized dissimilarity function $\sigma'$ defined as $\sigma'(i,j)=\sigma(i,j) + \sigma(j,i)$.

\subsection{Two-dimensional Embedding and Families}

To produce a visualization of the landscape of generators according to the above dissimilarity measure, we model these dissimilarities as distances in geometric space. We apply a metric \emph{Multi-Dimensional Scaling}~\citep{borg2005modern} algorithm\footnote{We used the metric MDS manifold embedding provided by the \emph{scikit-learn} Python module.} (MDS) to map them into a two-dimensional space. Distances between pairs of points are set to match dissimilarity values as closely as possible.

\begin{figure}
\begin{center}
\includegraphics[width=.8\linewidth]{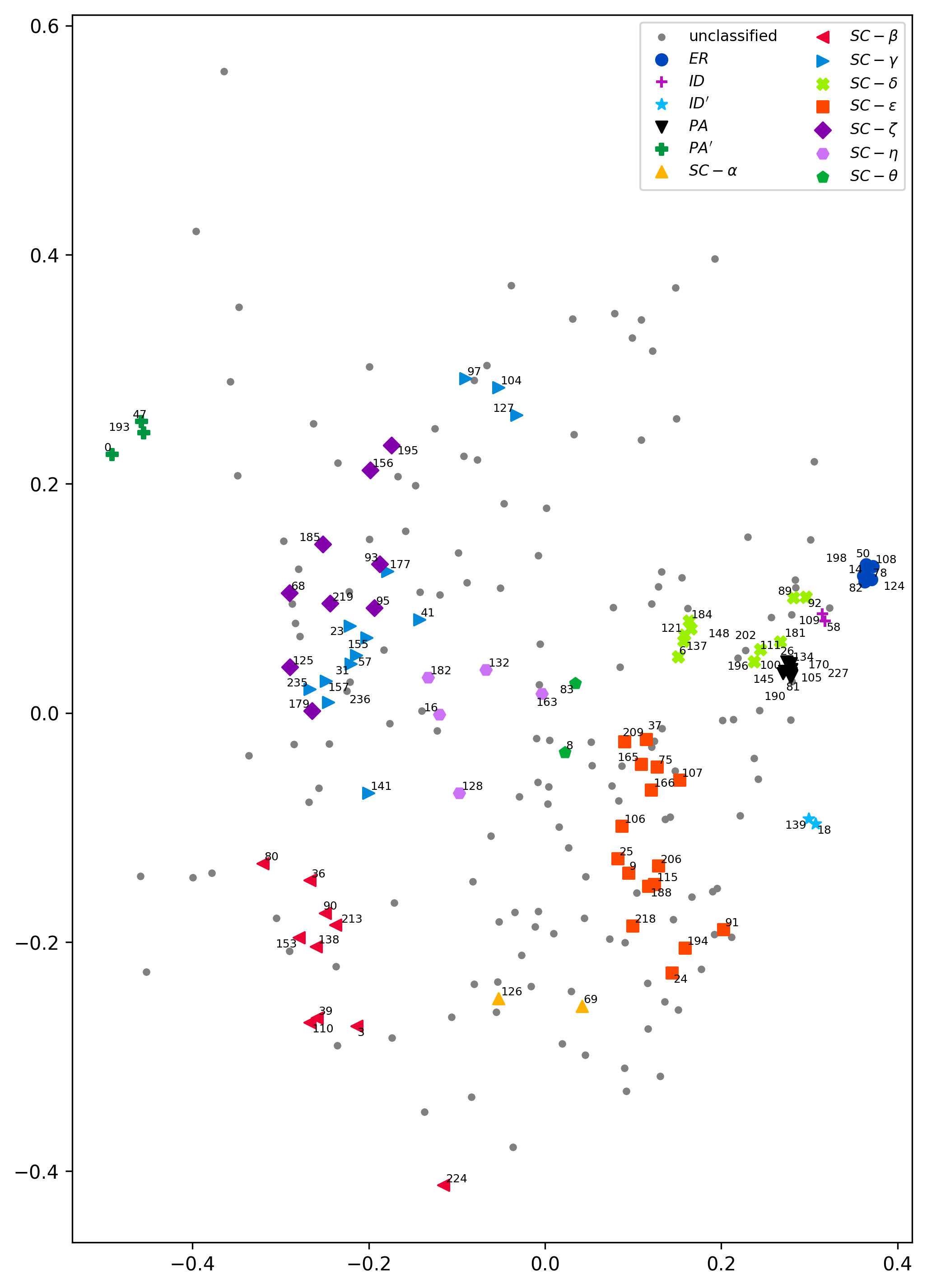}
\end{center}
\caption{Network generators mapped into a two-dimensional layout according to their pairwise distances. Different colors and shapes indicate families of generators that were manually identified as semantically similar. The legend shows the pattern that identifies each family.}
\label{fig:generators}
\end{figure}

We present the result of this two-dimensional embedding in figure~\ref{fig:generators}. 

\medskip
We also performed a manual analysis, looking for patterns of similar generators in mathematical terms, \hbox{i.e.} at the level of the explicit formula. We identified $11$ such strong patterns, and labeled every generator that conforms to one of them. We refer to sets of generators that conform to such patterns as \emph{families} ($n'=91$). The other ones are described as \emph{unclassified} ($n''=147$). 

This manual classification is presented on table~\ref{tab:families}, along with the actual generators assigned to each family.\footnote{Given the undirected nature of these networks, we simplify the notation for generators that use only variables from either the origin side or target side. Suppose we have the generator $3.k_i + d$; here, this is equivalent to $3.k_j + d$, so we simply write $3.k + d$.}

\medskip

The legend of figure~\ref{fig:generators} shows names that we gave to each family in table~\ref{tab:families}, based on their main common primary mathematical features (we detail the meaning of these names below). A first interesting observation of this result is that the families are distributed in spatial clusters. Visual inspection makes it quite clear that mathematically close generators appear in similar regions of the 2D plane, some being much more spread than others. Another interesting point is that, although many generators are left unclassified, families are spread across most of the extent of the overall spatial distribution.

\newcommand{\ER}{{\bf ER}}
\newcommand{\ID}{{\bf ID}}
\newcommand{\IDp}{{\bf ID'}}
\newcommand{\PA}{{\bf PA}}
\newcommand{\PAp}{{\bf PA'}}

\newcommand{\GID}[1]{\texttt{\tiny $\langle$#1$\rangle$}}
\begin{table}
\begin{scriptsize}
\begin{center}
\begin{tabular}{p{7.0mm}p{1mm}p{1mm}p{15.0mm}p{15.0mm}p{15.0mm}p{15.0mm}p{15.0mm}p{15.0mm}}
\multicolumn{3}{l}{\textbf{\emph{Family}}}&\multicolumn{6}{l}{\bf\emph{List of generator functions and corresponding network number} \rm\texttt{\scriptsize$\langle$ID$\rangle$}}\\
  \toprule

                                 &&& $ 0.08 $ & $ 0.88 $ & $ 0.95 $ & $ 54.6 $ & $ 0.62 $ & $ 6.0 $ \\
  \multirow{1}{*}{\textbf{ER}}   &&& \GID{14} & \GID{50} &  \GID{78} & \GID{82} & \GID{108} & \GID{124}  \\
  \multirow{1}{*}{$c$}           &&& \multicolumn{3}{l}{$ (\max(k_i, i)=0 \to 0, 0.63) $}\\
                                 &&& \GID{198} \\   
  \midrule

  \multirow{1}{*}{\textbf{ID}}   &&& $ i $ & $ i $\\
  \multirow{1}{*}{$i$}           &&& \GID{58} & \GID{109} \\ 
  \midrule

  \multirow{1}{*}{\textbf{ID'}}  &&& $e^i$ & $e^i$\\
  \multirow{1}{*}{$e^i$}         &&& \GID{18} & \GID{139} \\
  \midrule

                                 &&& $ k $ & $ k $ & $ k $ & $ k $ & $ k $ & $ k $ \\
  \multirow{1}{*}{\textbf{PA}}   &&& \GID{26} & \GID{81} & \GID{100} & \GID{105} & \GID{111} & \GID{134} \\
  \multirow{1}{*}{$k$}           &&& $ k $ & $ k $ & $ k $ \\
                                 &&& \GID{145} & \GID{170} & \GID{227} \\
  \midrule

  \multirow{1}{*}{\textbf{PA'}}  &&& ${k_j}^{k_i}$ & \multicolumn{4}{l}{$ (\min(j, .66) > k_i \to j, e^{k_j})^{(\min((j=0, k_j, k_i), e^{k_j}))}$} &${k_i}^{k_j}$\\
  \multirow{1}{*}{$k_i^{k_j}$}   &&& \GID{0} & \multicolumn{4}{l}{\GID{47}} & \GID{193}\\

  \midrule

  \multirow{1}{*}{\textbf{SC-$\alpha$}} &&& \multicolumn{2}{l}{$\psi_8(k_j^2, .62) - k_i$} & $\psi_7(k^3, 4)$ \\
  \multirow{1}{*}{$\psi_g(k^s, c)$}     &&& \multicolumn{2}{l}{\GID{69}} & \GID{126} \\

  \midrule

  &&& \multicolumn{1}{l}{$\psi_3(2^k, .48)$} & \multicolumn{1}{l}{$\psi_9(e^{k_i}, .49)$} & \multicolumn{1}{l}{$\psi_4(e^k, 1.1)$} & \multicolumn{2}{l}{$\psi_5(\frac{e^{\max(k_i, k_j)}}{k_i}, k_i)$} & \multicolumn{1}{l}{$\psi_5(e^k, 1)$} \\
  \multirow{1}{*}{\textbf{SC-$\beta$}}  &&& \multicolumn{1}{l}{\GID{3}} & \multicolumn{1}{l}{\GID{36}} & \multicolumn{1}{l}{\GID{39}} & \multicolumn{2}{l}{\GID{80}} & \multicolumn{1}{l}{\GID{90}} \\
  \multirow{1}{*}{$\psi_g(e^k, > \frac{1}{2})$}   &&& \multicolumn{1}{l}{$\psi_4(e^k, 1)$} & \multicolumn{1}{l}{$\psi_8(e^k, d)$} & \multicolumn{1}{l}{$\psi_4(k, .67)^k$} & \multicolumn{1}{l}{$\psi_5(e^k, 1.7)$} & \multicolumn{1}{l}{$\psi_3(e^k, 2)$} \\
  &&& \multicolumn{1}{l}{\GID{110}} & \multicolumn{1}{l}{\GID{138}} & \multicolumn{1}{l}{\GID{153}} & \multicolumn{1}{l}{\GID{213}} & \multicolumn{1}{l}{\GID{224}} \\

  \midrule

  &&& \multicolumn{1}{l}{$\psi_9(k^k, 0)$} & \multicolumn{1}{l}{$\psi_6(3^k, 0)$} & \multicolumn{1}{l}{$\psi_4(4 \cdot k^5, 0)$} & \multicolumn{1}{l}{$\psi_8(k^k, 0)$} & \multicolumn{2}{l}{$\psi_3(e^{k_i + k_j}, .05)$}\\
  &&& \multicolumn{1}{l}{\GID{23}} & \multicolumn{1}{l}{\GID{31}} & \multicolumn{1}{l}{\GID{41}} & \multicolumn{1}{l}{\GID{57}} & \multicolumn{2}{l}{\GID{97}} \\
  \multirow{1}{*}{\textbf{SC-$\gamma$}} &&& \multicolumn{1}{l}{$\psi_3(e^k, 0)$} & \multicolumn{1}{l}{$\psi_3(2^k, 0)$} & \multicolumn{2}{l}{$\psi_6(e^{\psi_5(1, k)}, 0) + .07$} & \multicolumn{1}{l}{$\psi_7(e^k, 0)$} & \multicolumn{1}{l}{$\psi_4(e^k, .06)$} \\
  \multirow{1}{*}{$\psi_g(k^B, \sim 0)$} &&& \multicolumn{1}{l}{\GID{104}} & \multicolumn{1}{l}{\GID{127}} & \multicolumn{2}{l}{\GID{141}} & \multicolumn{1}{l}{\GID{155}} & \multicolumn{1}{l}{\GID{157}} \\
  &&& \multicolumn{1}{l}{$\psi_2(k_i \cdot e^{k_j}, 0)$} & \multicolumn{1}{l}{$\psi_4(e^k, 0)$} & \multicolumn{1}{l}{$\psi_5(k^7, .01)$} & \multicolumn{1}{l}{$\psi_5(e^k, .03)$} \\
  &&& \multicolumn{1}{l}{\GID{164}} & \multicolumn{1}{l}{\GID{177}} & \multicolumn{1}{l}{\GID{235}} & \multicolumn{1}{l}{\GID{236}} \\
  
  \midrule
  
  &&& \multicolumn{1}{l}{$\psi_4(e^i, e^{k_i})$} & \multicolumn{1}{l}{$\psi_4(i^j, k_j)$} & \multicolumn{1}{l}{$\psi_2(j^i, k_i)$} & \multicolumn{1}{l}{$\psi_3(e^i, k_i)$} & \multicolumn{1}{l}{$\psi_3(e^i, e^7)$} & \multicolumn{1}{l}{$\psi_3(e^i, 1)$} \\
  \multirow{1}{*}{\textbf{SC-$\delta$}} &&& \multicolumn{1}{l}{\GID{6}} & \multicolumn{1}{l}{\GID{89}} & \multicolumn{1}{l}{\GID{92}} & \multicolumn{1}{l}{\GID{121}} & \multicolumn{1}{l}{\GID{137}} & \multicolumn{1}{l}{\GID{148}} \\
  \multirow{1}{*}{$\psi_g(e^i, *)$} &&& \multicolumn{1}{l}{$\psi_2(9^i, 9^9)$} & \multicolumn{1}{l}{$\psi_3(e^i, j)$} & \multicolumn{2}{l}{$\psi_3(e^{i + j - d}, e^5)$} & \multicolumn{1}{l}{$\psi_4(9^i, 9)$} \\
  &&& \multicolumn{1}{l}{\GID{181}} & \multicolumn{1}{l}{\GID{184}} & \multicolumn{2}{l}{\GID{196}} & \multicolumn{2}{l}{\GID{202}} \\

  \midrule
  
  &&& \multicolumn{1}{l}{$9 \psi_3(i k_i, 2k_i)$} & \multicolumn{1}{l}{$\psi_4(i k_j, 6k_j)$} & \multicolumn{1}{l}{$\psi_5(ik_i, k_i)$} & \multicolumn{1}{l}{$\psi_9(ik_i, .1k_i)$} & \multicolumn{1}{l}{$\psi_2(ik_i, k_i)$} & \multicolumn{1}{l}{$\psi_7(ik_i, 7k_i)$} \\
  &&& \multicolumn{1}{l}{\GID{9}} & \multicolumn{1}{l}{\GID{24}} & \multicolumn{1}{l}{\GID{25}} & \multicolumn{1}{l}{\GID{37}} & \multicolumn{1}{l}{\GID{75}} & \multicolumn{1}{l}{\GID{91}} \\
  \multirow{1}{*}{\textbf{SC-$\epsilon$}} &&& \multicolumn{1}{l}{$\psi_6(ik_i, .44k_i)$} & \multicolumn{1}{l}{$\psi_4(jk_i, .38)$} & \multicolumn{1}{l}{$\psi_3(jk_i, k_j)$} & \multicolumn{2}{l}{$\psi_4(i\log(k_i), 0)$} & \multicolumn{1}{l}{$\psi_3(jk_i, \frac{k_i}{4})$} \\
  \multirow{1}{*}{$\psi_g(i k, *)$} &&& \multicolumn{1}{l}{\GID{106}} & \multicolumn{1}{l}{\GID{107}} & \multicolumn{1}{l}{\GID{115}} & \multicolumn{2}{l}{\GID{165}} & \multicolumn{1}{l}{\GID{166}}\\
  &&& \multicolumn{2}{l}{$(\frac{k_j k_i}{.66} + d) \psi_4(j, .61)$} & \multicolumn{1}{l}{$\psi_3(i k_j, 2 k_j)$} & \multicolumn{1}{l}{$\psi_3(i k_j, k_j)$} & \multicolumn{1}{l}{$\psi_3(i k_i, 0)$} & \multicolumn{1}{l}{$\psi_4(i k_i, 3 k_i)$} \\
  &&& \multicolumn{2}{l}{\GID{188}} & \multicolumn{1}{l}{\GID{194}} & \multicolumn{1}{l}{\GID{206}} & \multicolumn{1}{l}{\GID{209}} & \multicolumn{1}{l}{\GID{218}} \\

  \midrule
  
  &&& \multicolumn{1}{l}{$\psi_7(i, 0)^{k_j}$} & \multicolumn{1}{l}{$\frac{7}{d} \psi_4(i^{k_i}, .48)$} & \multicolumn{1}{l}{$\psi_4(\frac{i^{k_j}}{k_j}, .18)$} & \multicolumn{1}{l}{$\psi_8(i^{k_i}, 2)$} & \multicolumn{1}{l}{$\psi_4(i^{k_i}, 0)$} & \multicolumn{1}{l}{$\psi_4(\frac{1}{6} i^{k_i}, d)$} \\
  \multirow{1}{*}{\textbf{SC-$\zeta$}} &&& \multicolumn{1}{l}{\GID{68}} & \multicolumn{1}{l}{\GID{93}} & \multicolumn{1}{l}{\GID{95}} & \multicolumn{1}{l}{\GID{125}} & \multicolumn{1}{l}{\GID{156}} & \multicolumn{1}{l}{\GID{179}} \\
  \multirow{1}{*}{$\psi_g(i^k, *)$} &&& \multicolumn{1}{l}{$\psi_9(d j^{k_i}, 0)$} & \multicolumn{2}{l}{$\psi_{\min(i, 4)}(i^{k_i}, 0)$} & \multicolumn{1}{l}{$\psi_5(9j^{k_i}, .03)$} \\
  &&& \multicolumn{1}{l}{\GID{185}} & \multicolumn{2}{l}{\GID{195}} & \multicolumn{1}{l}{\GID{219}} \\

  \midrule
  
  &&& \multicolumn{1}{l}{$\psi_5((i k_i)^2, i)$} & \multicolumn{1}{l}{$\psi_5(i {k_i}^2, 6)$} & \multicolumn{2}{l}{$\psi_4(2980.96k^2, 2k)$} & \multicolumn{2}{l}{$\psi_2(i{k_j}^2, {k_j}^2)$} \\
  \multirow{1}{*}{\textbf{SC-$\eta$}} &&& \multicolumn{1}{l}{\GID{16}} & \multicolumn{1}{l}{\GID{128}} & \multicolumn{2}{l}{\GID{132}} & \multicolumn{2}{l}{\GID{163}} \\
  \multirow{1}{*}{$\psi_g(i k^2, *)$} &&& \multicolumn{2}{l}{$\psi_7(\psi_i(.5, {k_j}^2), 0)$} \\
  &&& \multicolumn{2}{l}{\GID{182}} \\
  
  \midrule
  
  \multirow{1}{*}{\textbf{SC-$\theta$}} &&& \multicolumn{2}{l}{$\psi_4(k, 0) - .99$} & \multicolumn{2}{l}{$\psi_7(k, 0) - .93 $} \\
  \multirow{1}{*}{$\psi_g(k, 0) - 1$} &&& \multicolumn{2}{l}{\GID{8}} & \multicolumn{2}{l}{\GID{83}} \\
  
  \bottomrule
\end{tabular}
\end{center}
\end{scriptsize}
\caption{Generator expressions for each family. $c$ represents a constant value, $s$ a small exponent, $B$ a big exponent and $*$ is used as a placeholder for an arbitrary expression.}
\label{tab:families}
\end{table}

\medskip
In the middle-right region of figure~\ref{fig:generators} we can find two families that correspond to well-known network models. The first is family \ER, of the generators that are defined by some constant value $c$. They assign the same probability to every potential edge, and thus correspond to \emph{Erd\H{o}s-R\'{e}nyi} random graphs. The second is family \PA, of the generators that are defined by the degree variable $k$. They assign to each potential edge a probability that is proportional to the degree of either the origin or the target, and thus correspond to pure \emph{Preferential Attachment} networks. It is interesting to observe that these two quintessential network formation explanations show up in our generator set, albeit in a small quantity. Further, they are relatively close to each other in respect to many other, more complex explanations. A third family of very simple generators (family \ID), is the one where the probability of a potential edge is proportional to the sequential identifier of either the origin or the target. These generators are defined by the expression $i$. These are the simplest possible generators that take into account non-topological or exogenous features of nodes. This family is situated between the \ER{} and \PA{} families. Two other families exhibit expressions which roughly appear to be exponential versions of \ID{} and \PA{}. We named them \IDp{} and \PAp{}: they  nonetheless behave very distinctly as the exponential induce a strong winner-takes-all effect on the highest value of the main variable ($i$ or $k$). They are also situated in parts of the space distinct of their linear counterparts. 

Notice that for these simple cases, although many of the generators are exactly the same, their positions do not coincide precisely in the spatial embedding. This is due to the fact that the generative process is stochastic, and some random variation is to be expected.

\newcommand{\SC}{{\bf SC}}
\newcommand{\SCm}[1]{{\bf SC-$#1$}}
The first five families are very simple. The other eight have a very strong resemblance with one another: they all use the affinity function, based on some constant number of affinity groups. This means that link dynamics is strongly influenced by the existence of a certain number of classes of nodes which likely matches underlying \emph{Social Circles}; we denote this family as \SC. A simple interpretation for this is indeed that ego networks are a sample of social groups that ego belongs to. For example: school friends, family, work colleagues and so on. It makes sense that these groups are much more densely connected within themselves than between them, as they correspond to separate social spheres. The affinity function provides a very straightforward way of generating this type of linking behavior. The constant number of groups present in the first parameter of affinity functions represents an estimation of the number of social groups that ego belongs to.
In our previous work~\citep{menezes2014symbolic} we included one Facebook ego network in the diverse set of networks used, and the generator found for it was also based on an affinity function. In fact, under the typology we present here, it would be classified as an \SC-{$\epsilon$} generator. From the biological, social and technological networks analyzed in that work, the Facebook ego network was the only one based on an affinity function with a constant number of groups. This presents us with additional empirical evidence that this is in fact a characteristic signature of ego-centered social networks. 

\newcommand{\mynetx}[2]{\raisebox{-.5\height}{\includegraphics[width=#1\linewidth]{figs/netfigs/#2.pdf}}}
\newcommand{\mynet}[1]{\mynetx{.1}{#1}}
\newcommand{\mynetlegend}[1]{{\footnotesize\em(#1)}}

\begin{table}\centering
\begin{tabular}{lcccccc}
\toprule
&ER \GID{198}
&PA \GID{190}
&ID \GID{109}
&SC-$\gamma$ \GID{97}
&SC-$\delta$ \GID{181}
&SC-$\eta$~\GID{128}
\\\midrule
\mynetlegend{Real}
&\mynetx{.1}{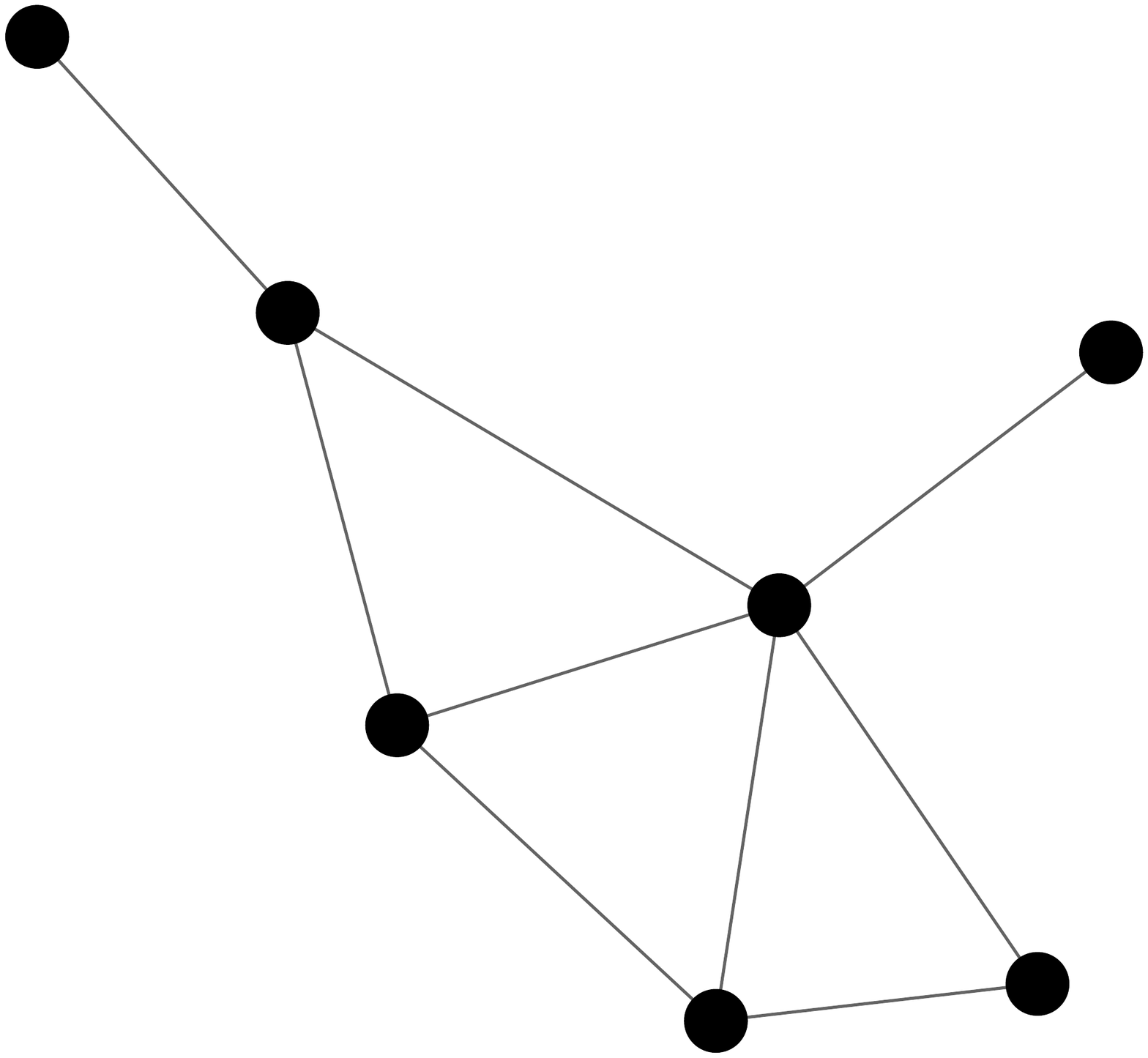}
&\mynetx{.12}{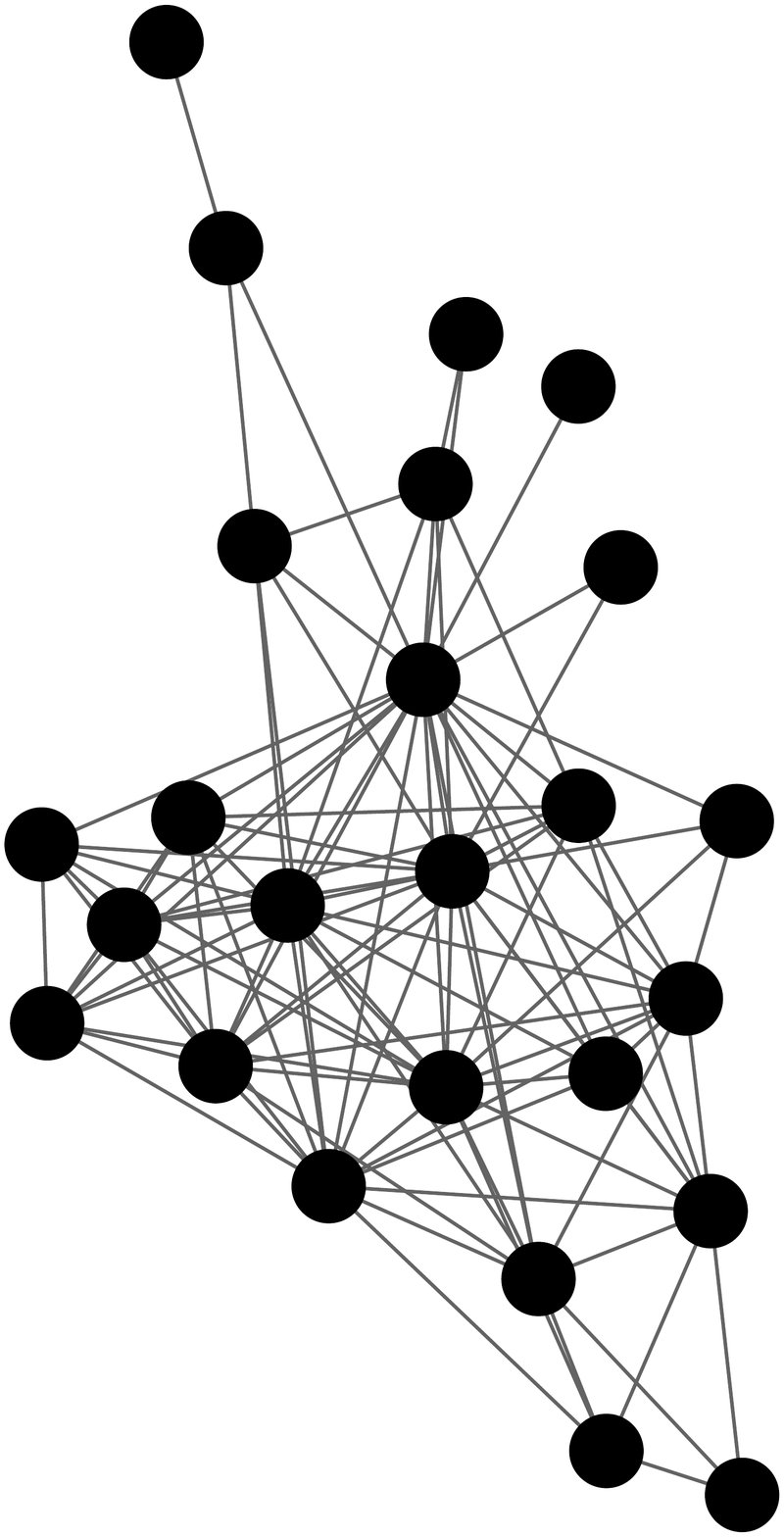}
&\mynet{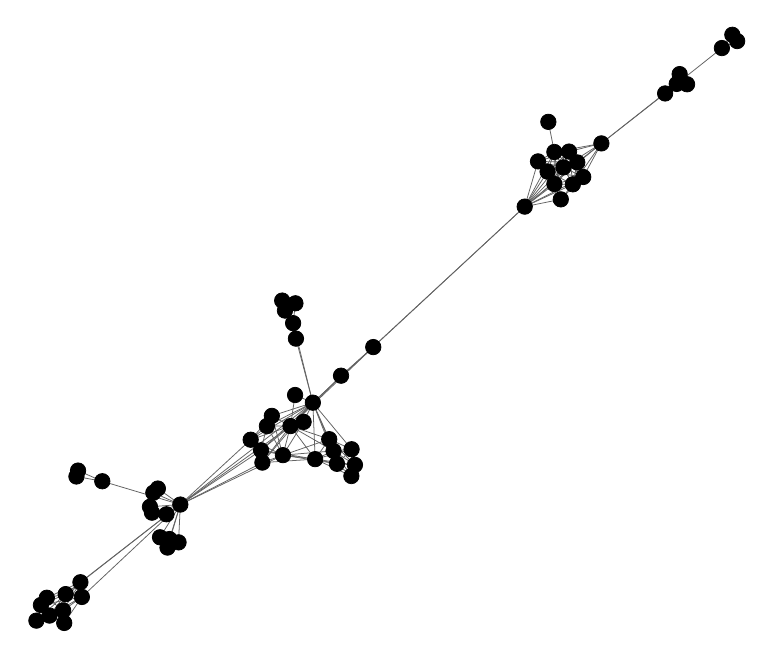}
&\mynet{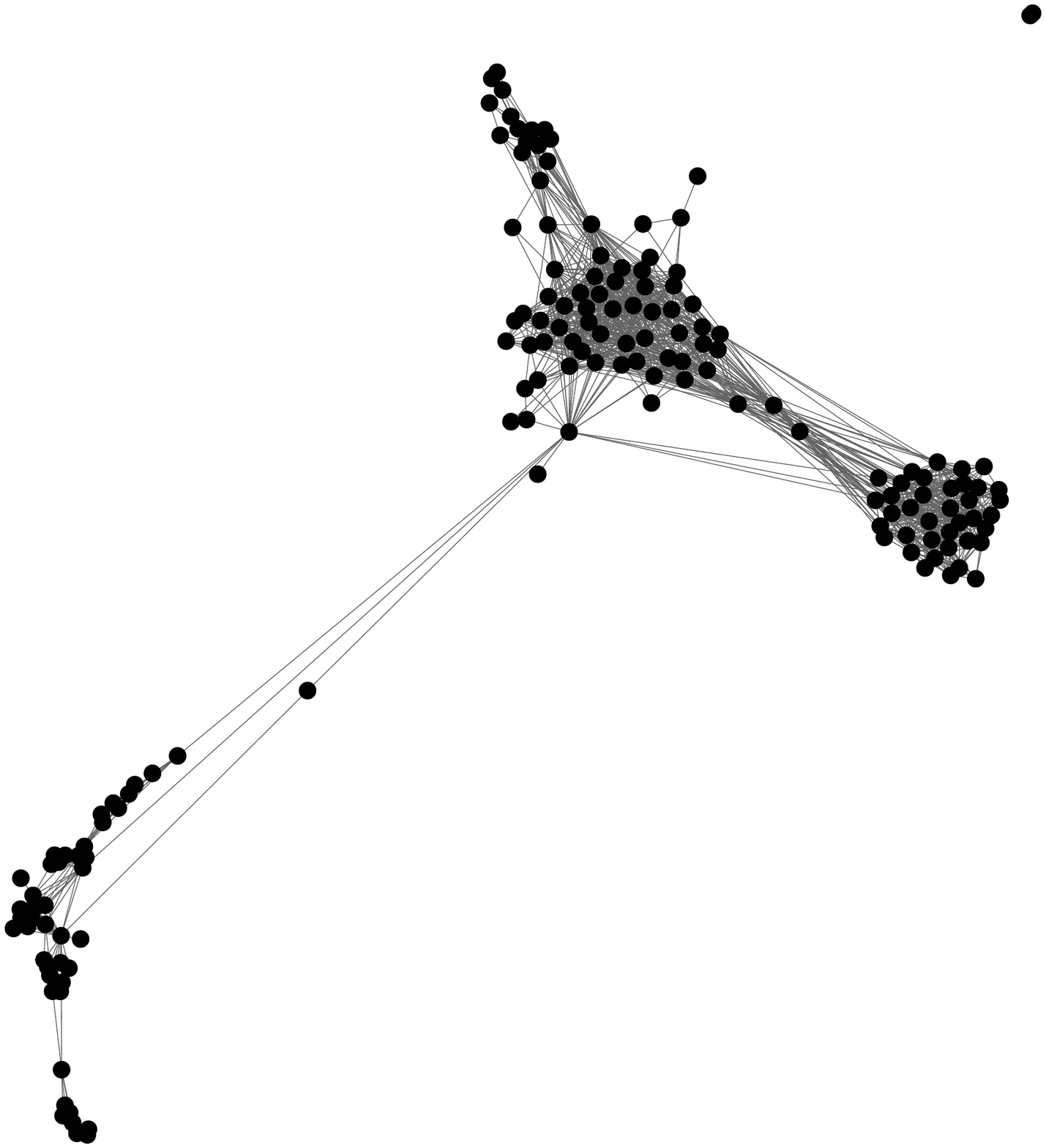}
&\mynet{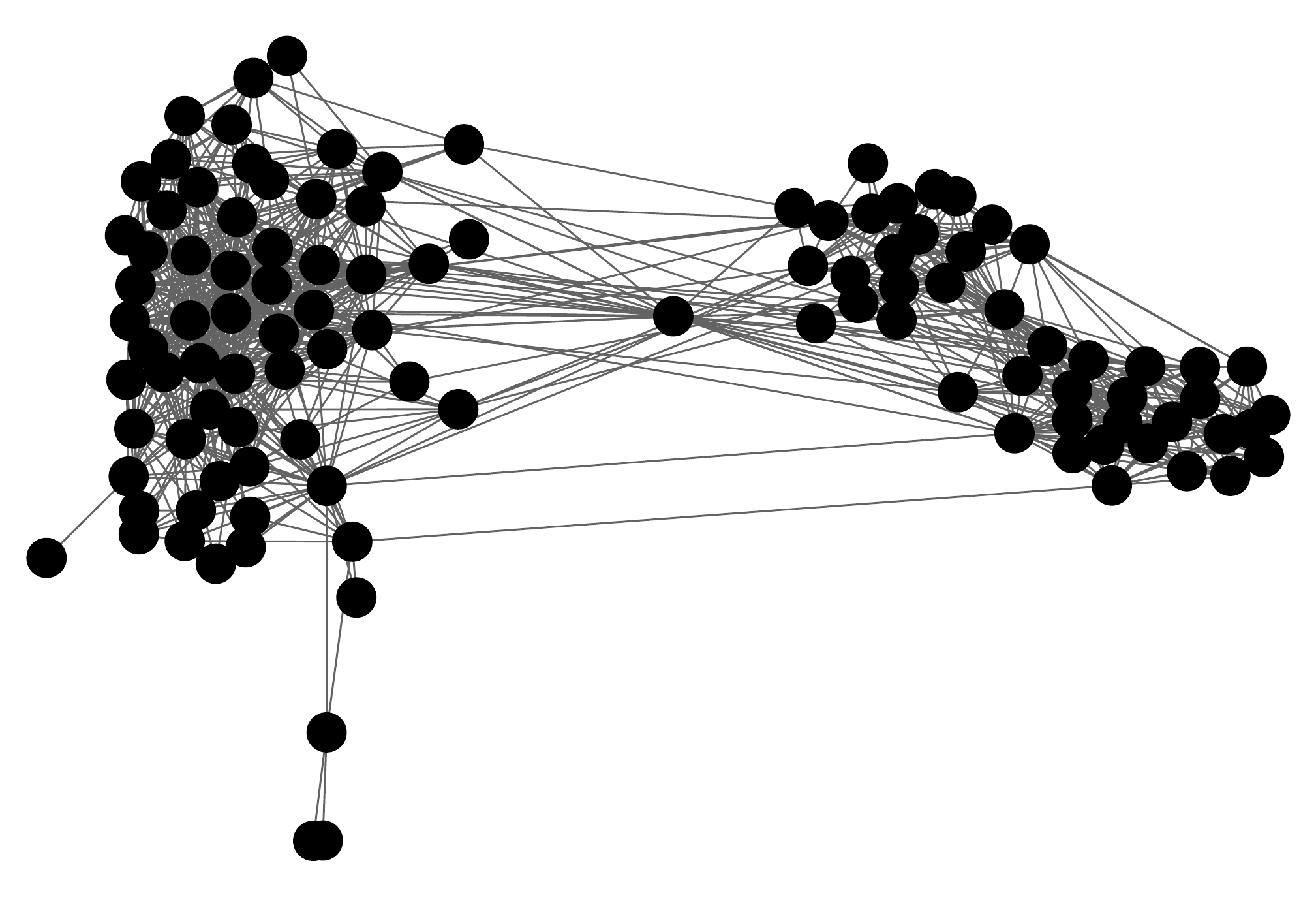}
&\mynet{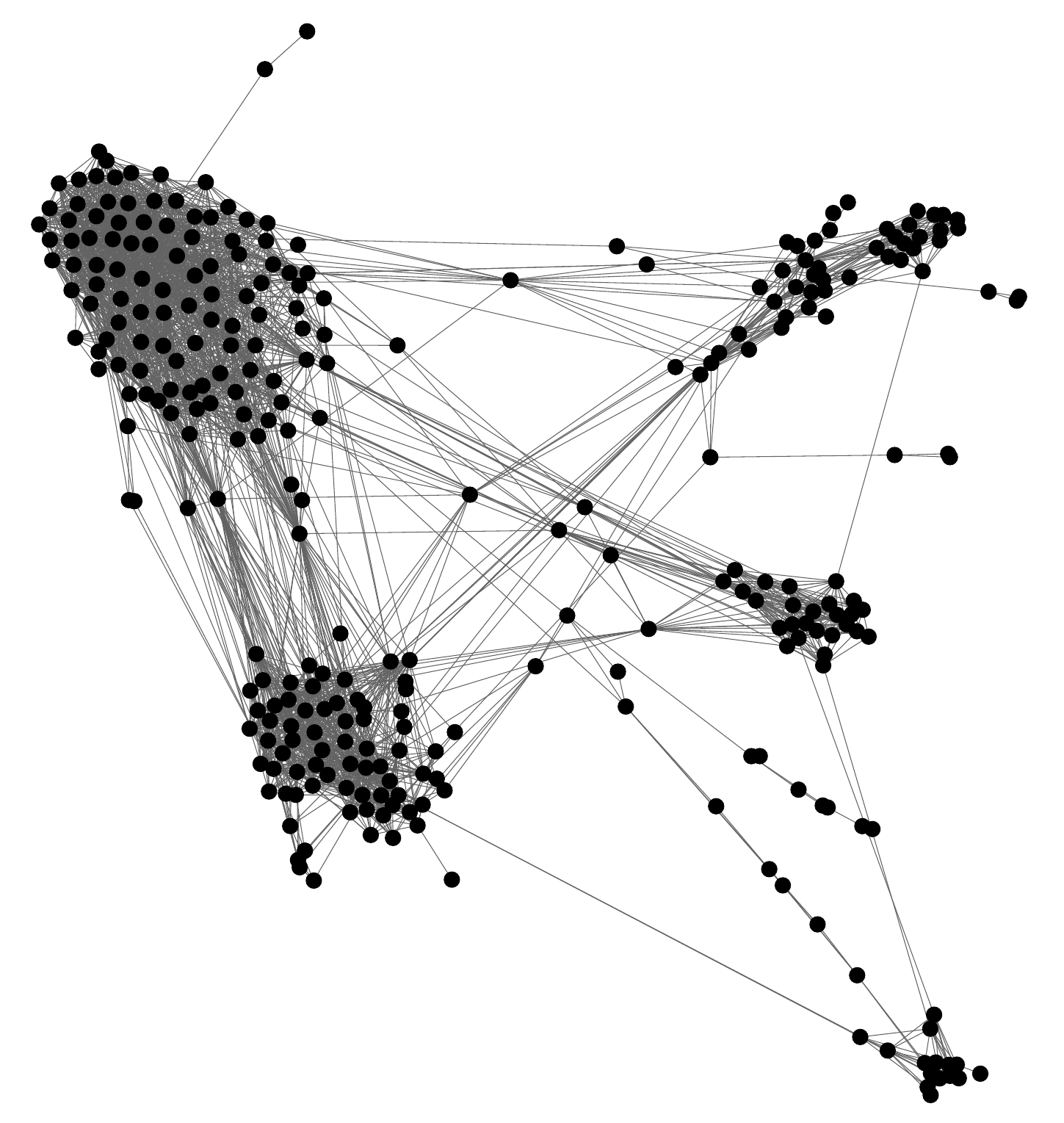}
\\
\mynetlegend{Synthetic}
&\mynetx{.1}{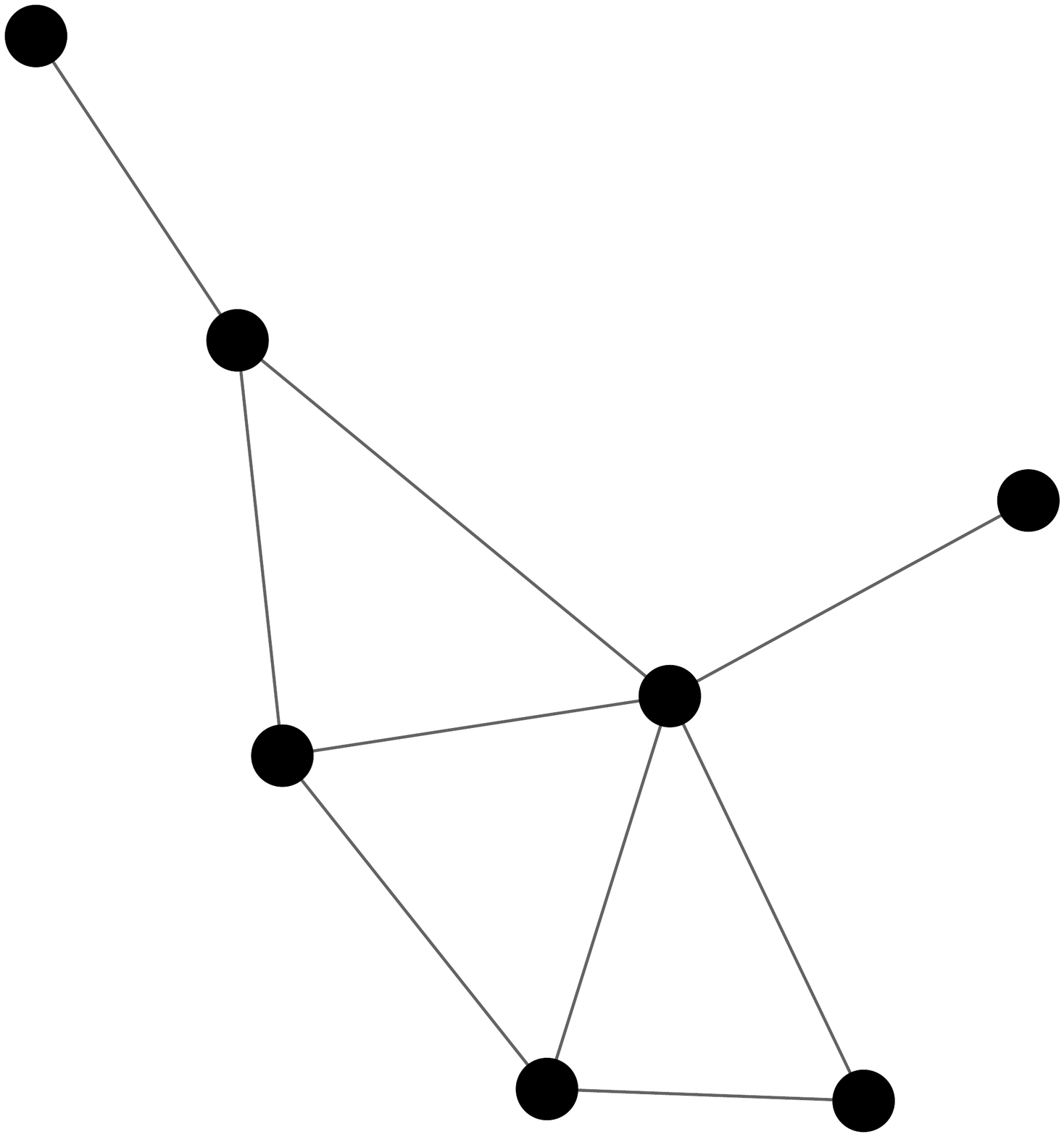}
&\mynetx{.12}{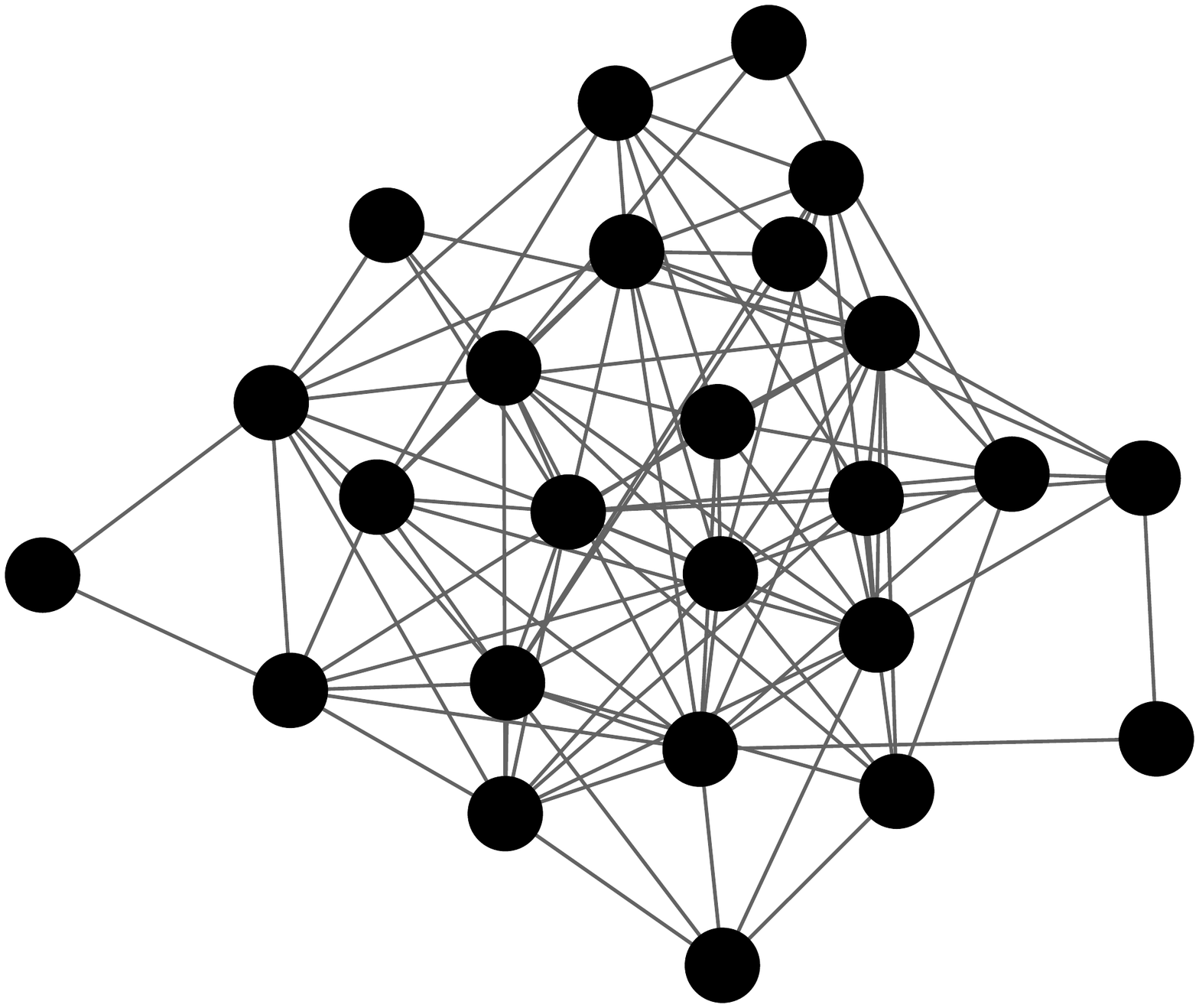}
&\mynet{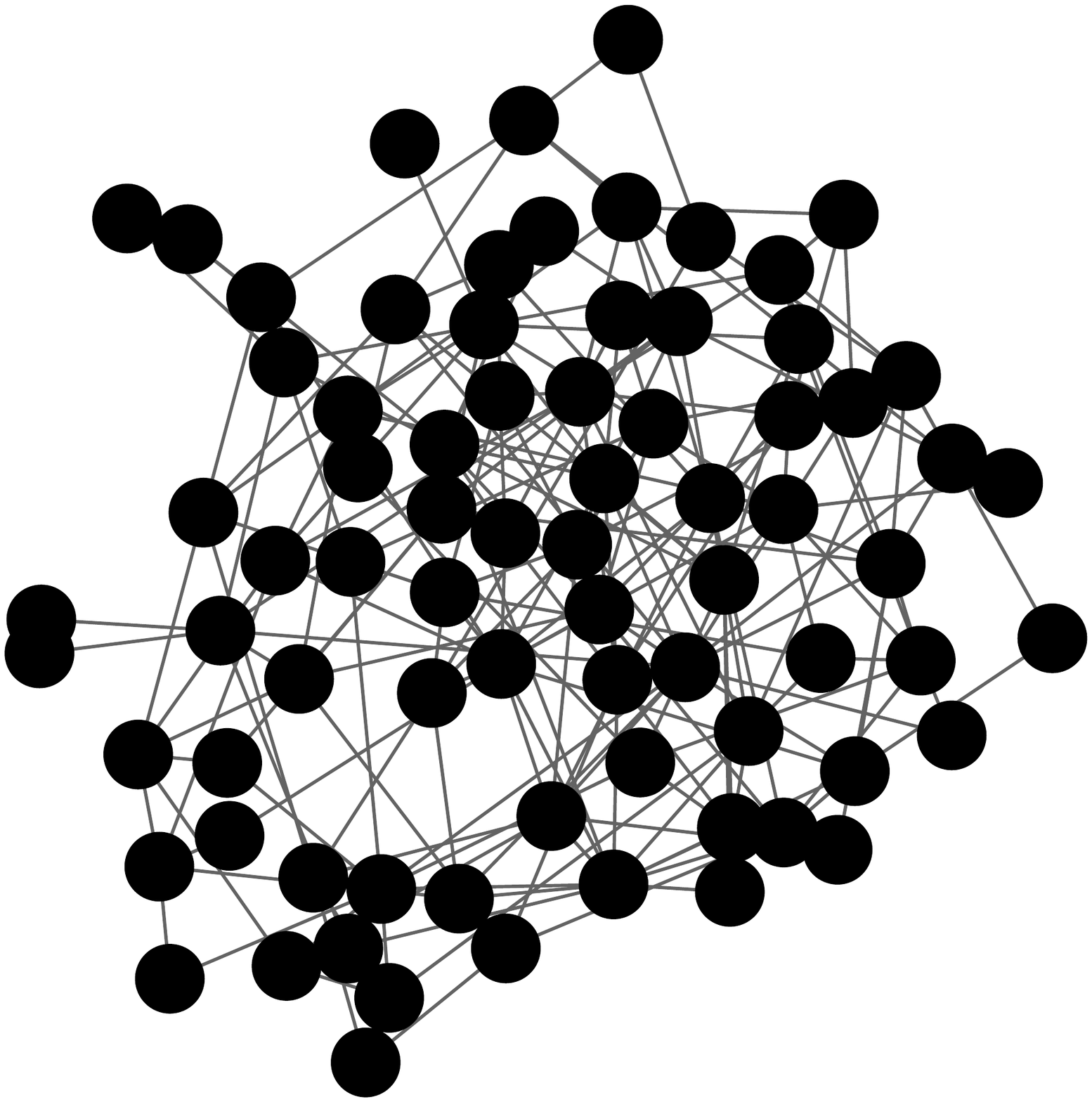}
&\mynet{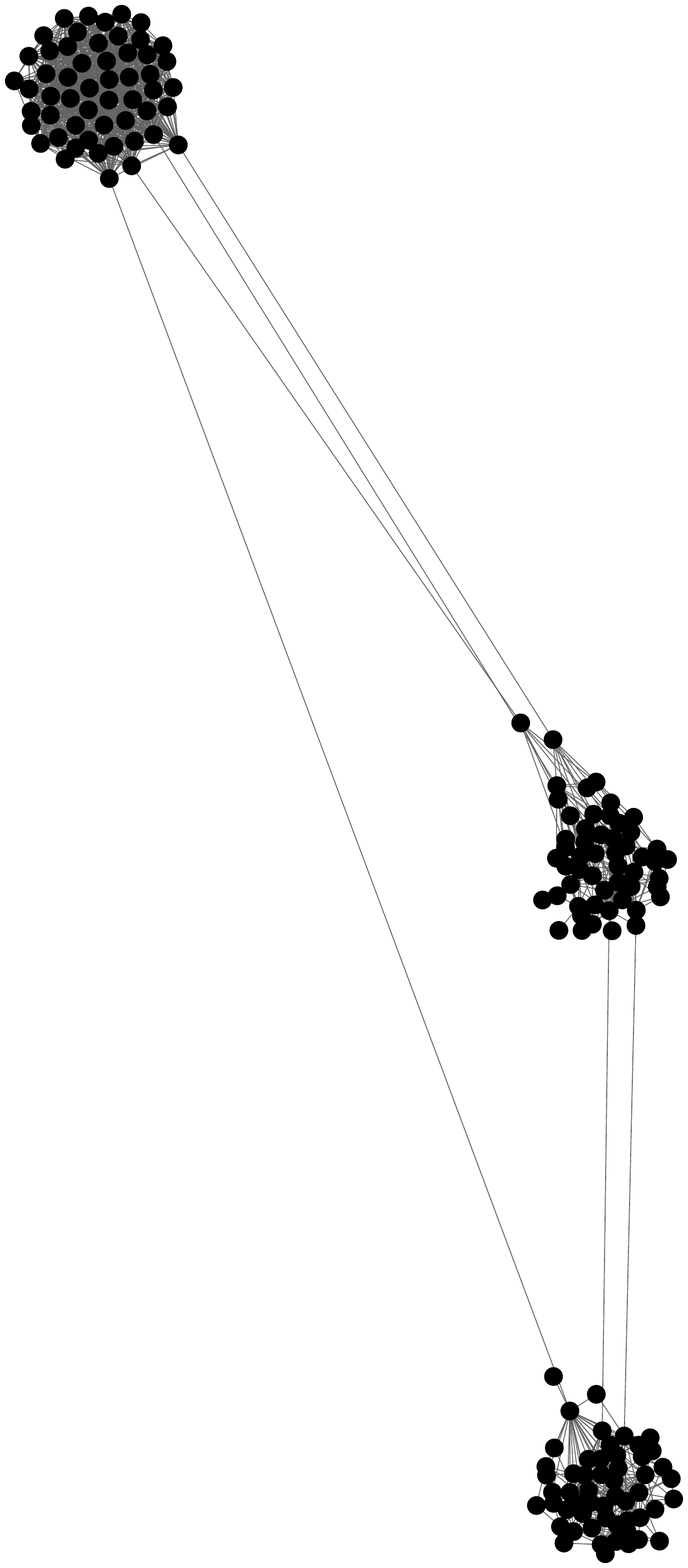}
&\mynet{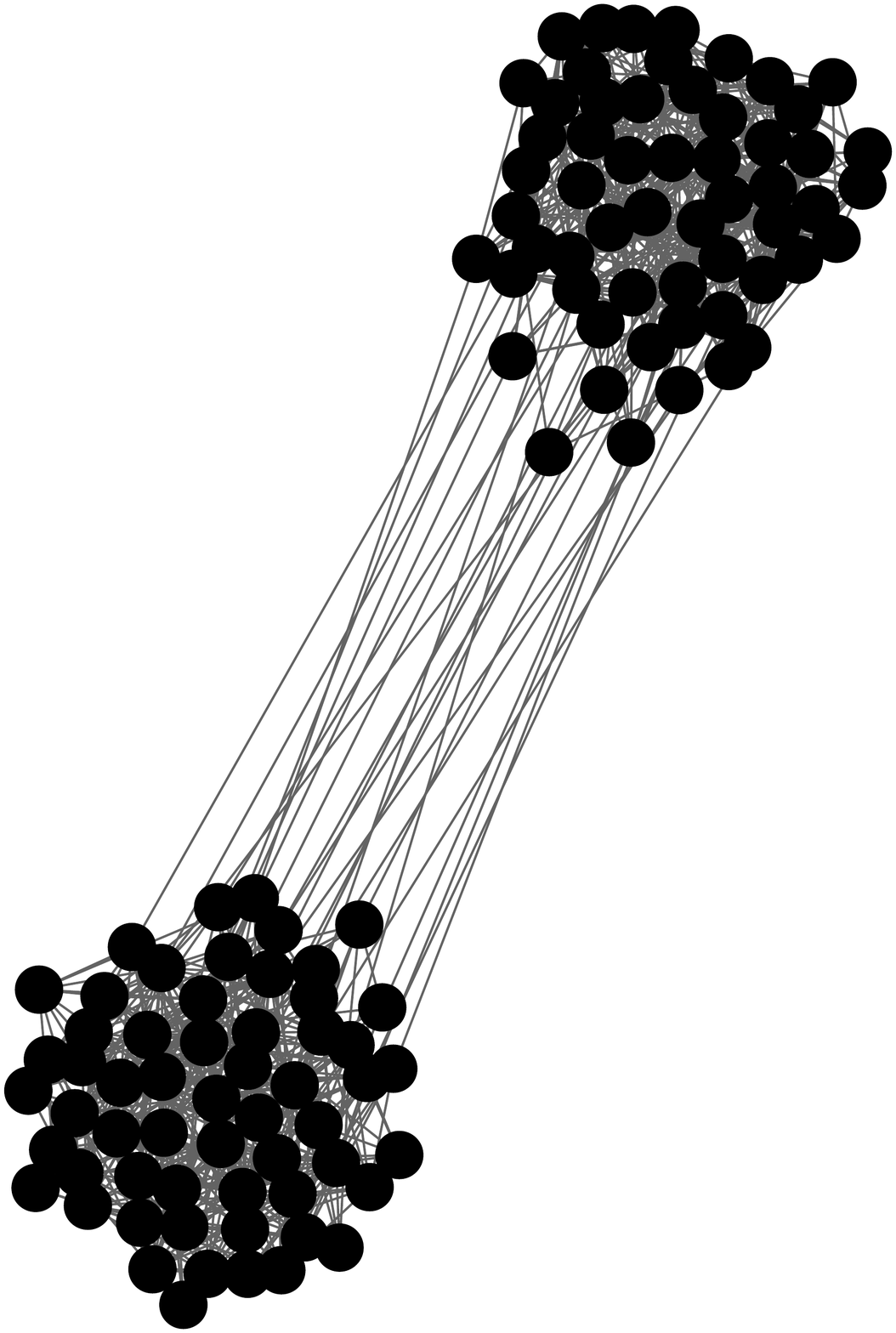}
&\mynet{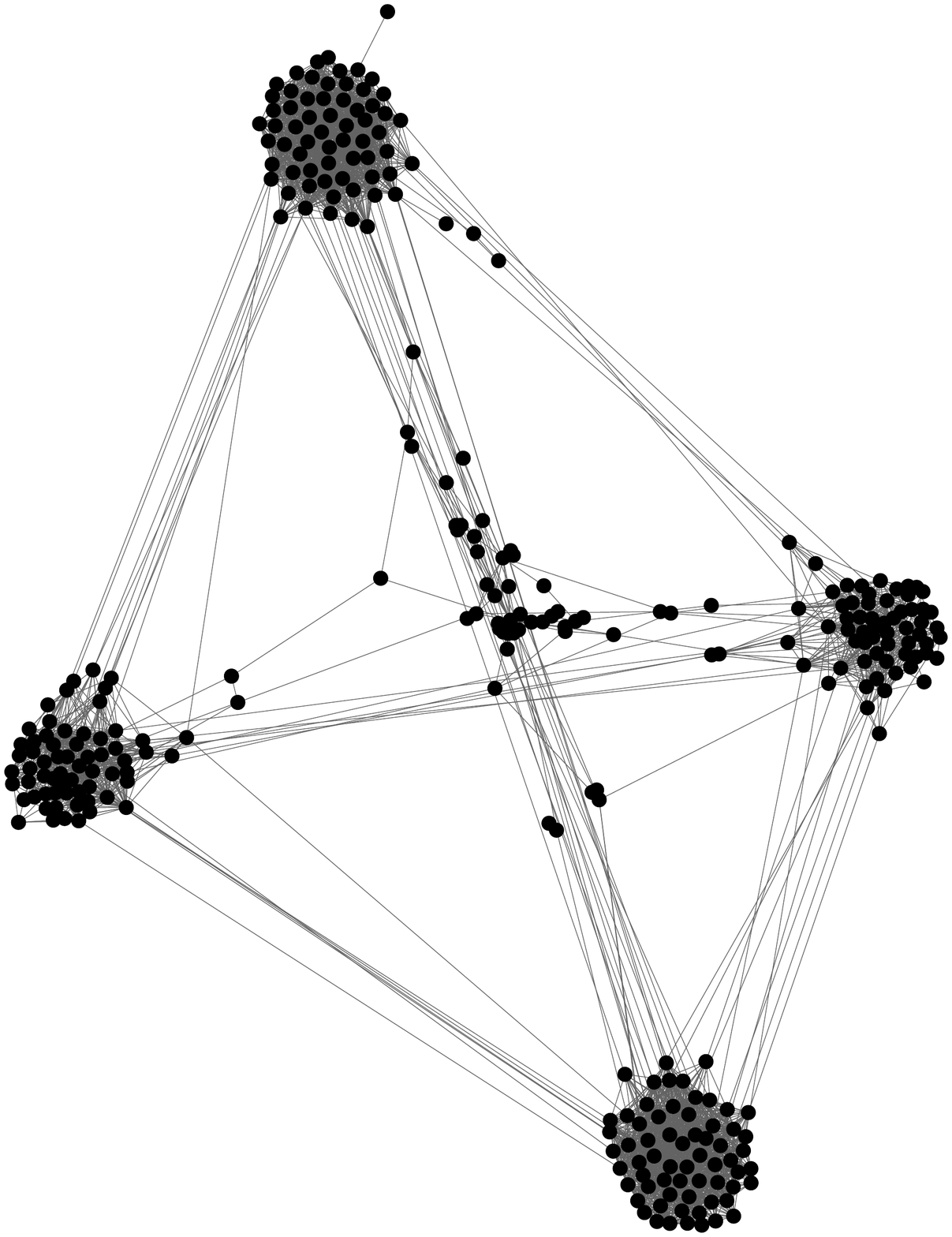}
\\\bottomrule
\end{tabular}
\caption{\label{tab:illustrations}Visual representation of some empirical ego-networks (top row) with their reconstruction (bottom row), for a selection of evoked families. ER, PA and ID are featured; each of the three main subfamilies of SC are also present (generators 97, 181 and 128 are all based on an affinity function of parameters 3, 2 and 5, respectively).
Note that three of the empirical networks (109, 128, 181) feature very small disconnected components, gathering no more than a handful of nodes which have not been drawn here for clarity purposes.}
\end{table}

To illustrate further these families, we provide a few visual examples of network generators on table~\ref{tab:illustrations}. For each selected generator of a given family, we put along the original empirical network and its reconstruction using the same number of nodes. Spatialization follows a force-directed layout. The number of social circles parameterized on $\psi$ may be seen to be faithful to the original number of clusters in the real network.

\SC{} families differ in the linking behavior for nodes deemed to belong to the same group. Some of them are purely based on topological factors (families $\alpha$, $\beta$, $\gamma$ and $\theta$), one only on exogenous factors (family $\delta$) and some on a combination of both (families $\epsilon$, $\zeta$ and $\eta$).

The largest family is $\epsilon$, which assigns probability of in-group links as a linear combination of current degree ($k$) and exogenous factors ($i$). The second largest family by number of generators found is family $\gamma$, and it is also the one that is the most spread in the spatial embedding. In this family, the probability of in-group connections is purely driven by topology, as an exponential of the current degree of one of the nodes. We can think of it as a form of super-preferential attachment within social circles -- current popularity within the group is highly rewarded. For most cases, the probability of connection between groups is given by a relatively small constant, and for a few it is zero.



\begin{figure}
\center\begin{tabular}{lp{.65cm}c}
\includegraphics[width=.35\linewidth]{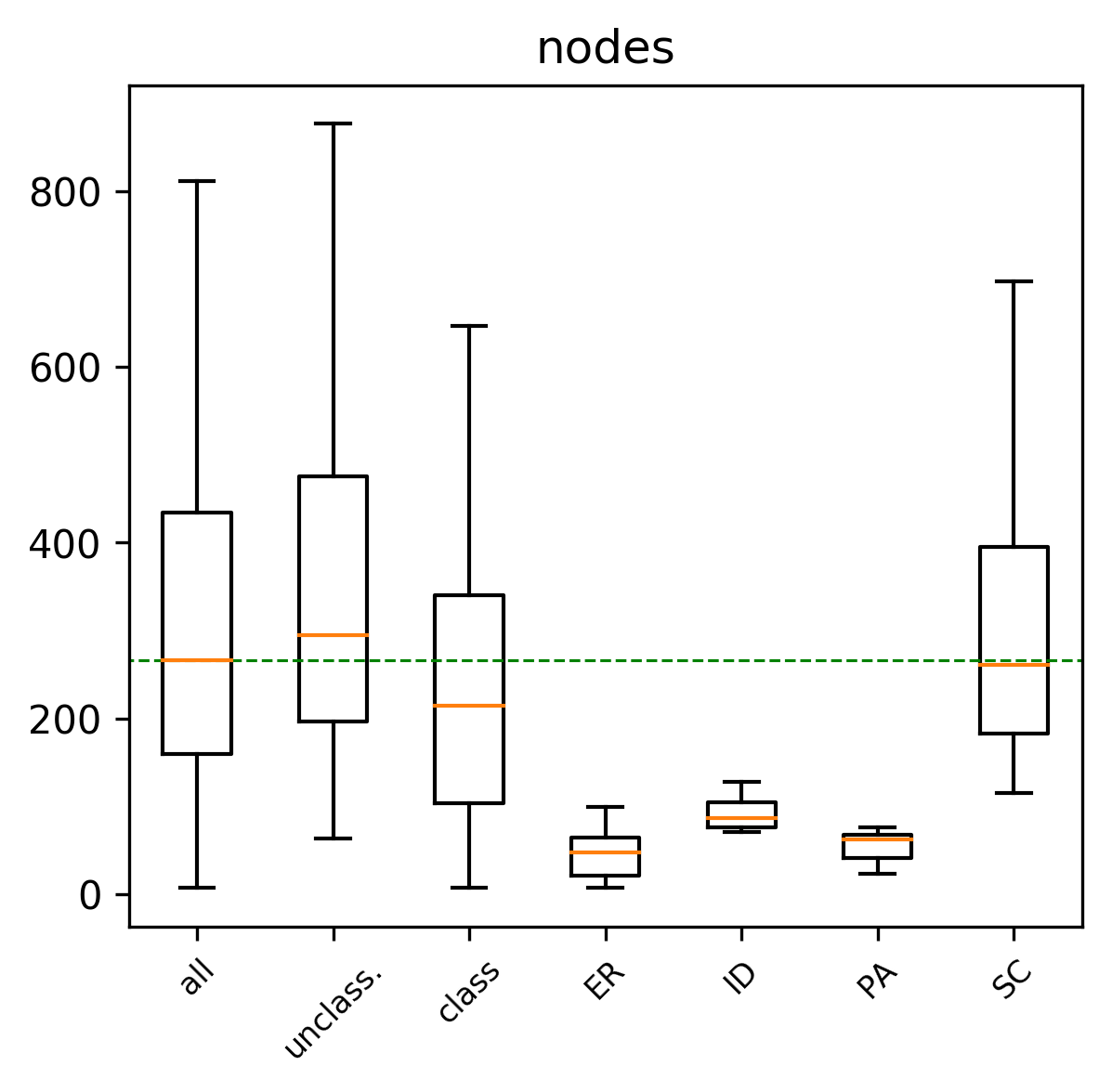}
&&
\includegraphics[width=.36\linewidth]{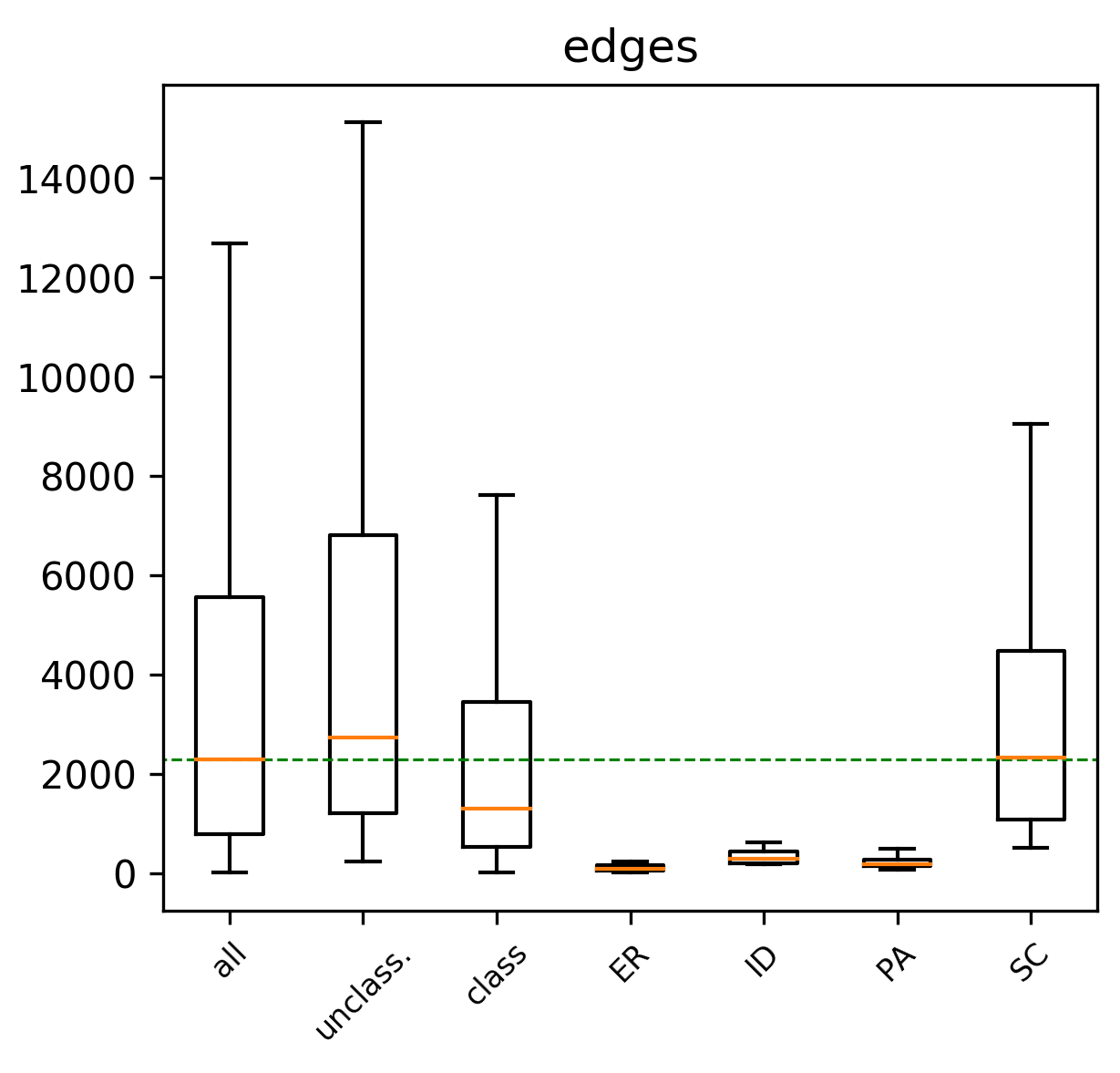}
\\
\includegraphics[width=.35\linewidth]{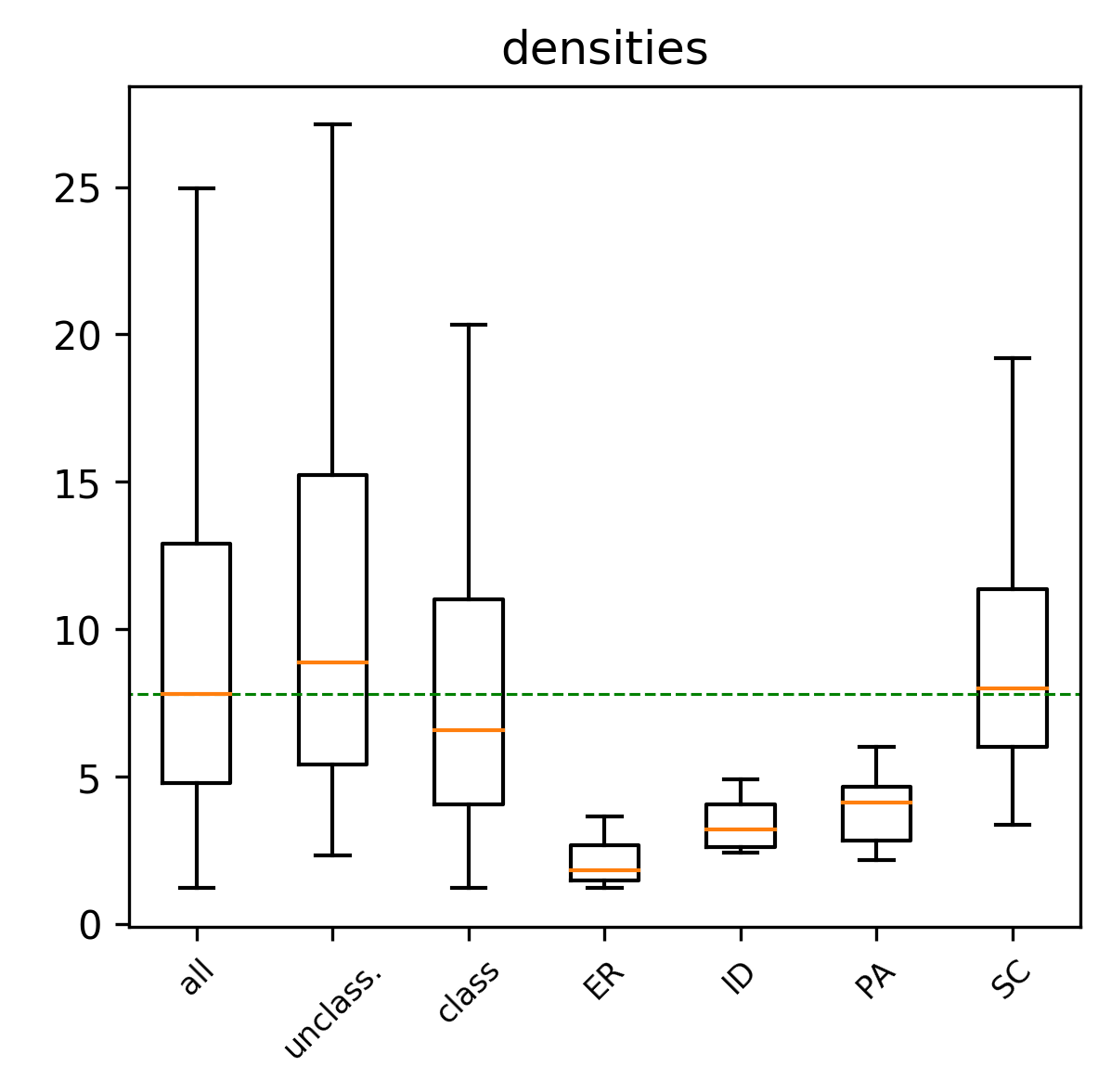}
&&
\includegraphics[width=.45\linewidth]{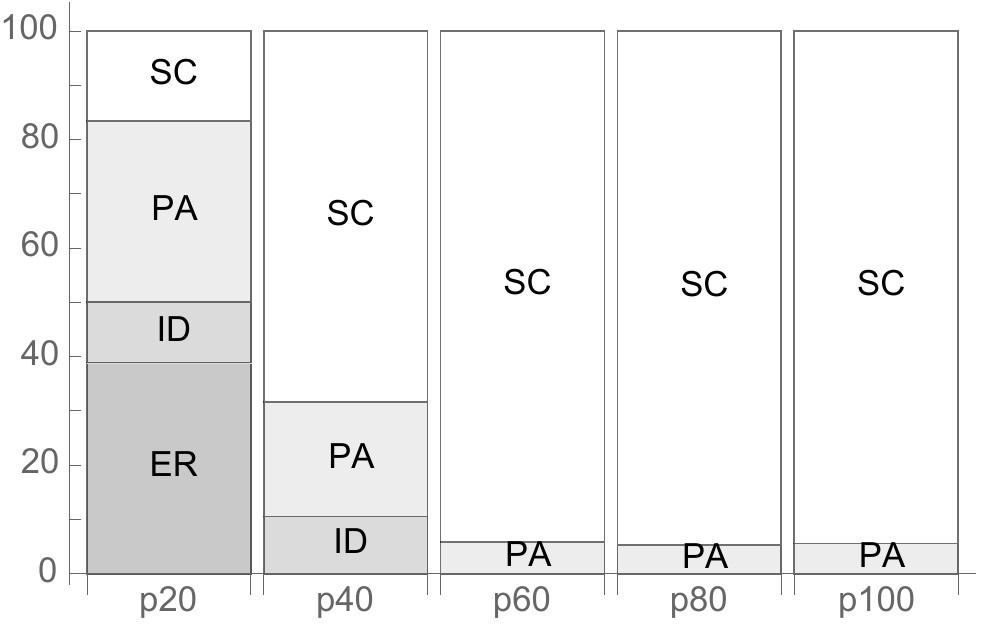}
\end{tabular}
\caption{{\em Top panel, and bottom-left:} Boxplots of numbers of nodes, edges and densities for the underlying networks of the various families, as well as all, unclassified and classified. Horizontal dashed line indicates overall median.
{\em Bottom-right:} Stacked plot of family ratio per percentile of network density.}
\label{fig:nodes_edges}\label{fig:percentiles_stacked}
\end{figure}

Some questions remain. Why are some generators so simple, and why are more than half of the generators so diverse that they cannot be classified into families? In an attempt to attain a better understanding, we created boxplots of the distributions of node and edge counts for the underlying networks per family, as well as for all generators, and for classified and unclassified generators. These plots are presented in figure~\ref{fig:nodes_edges}, as well as a stacked plot of family ratio per percentile of network density. Some interesting facts are revealed.

The families of simpler generators (\ER, \ID, \IDp, \PA{} and \PAp) have both node and edge counts well below the median. This could indicate that these simple generators correspond to cases where there is not enough data to form a more complex theory. The simplest underlying behavior is captured, corresponding precisely to the simple archetypal explanations of preferential attachment and random behavior. Maybe these networks are small because the corresponding user is not very active, or does not have many social connections, or maybe because they joined recently and the networks are at their initial stages of growth. When the latter case is true, our results seem to indicate that they may be assignable to a more complex family when they develop more. Under this assumption, families \SC{} paint here the more relevant part of the picture of network growth dynamics.

The unclassified set corresponds to networks that are slightly larger than the mean, both in numbers of nodes and edges. From this observation we formulate two hypotheses. The first one is that the unclassified set really does correspond to a complex variety of behaviors. It could be that, given one or two orders of magnitude more ego networks, more families would be found. The second one is that it is more difficult for evolutionary search to find simple generators for these larger networks, but that given more runs, they would emerge.

\begin{figure}
\centering
\includegraphics[width=.55\linewidth]{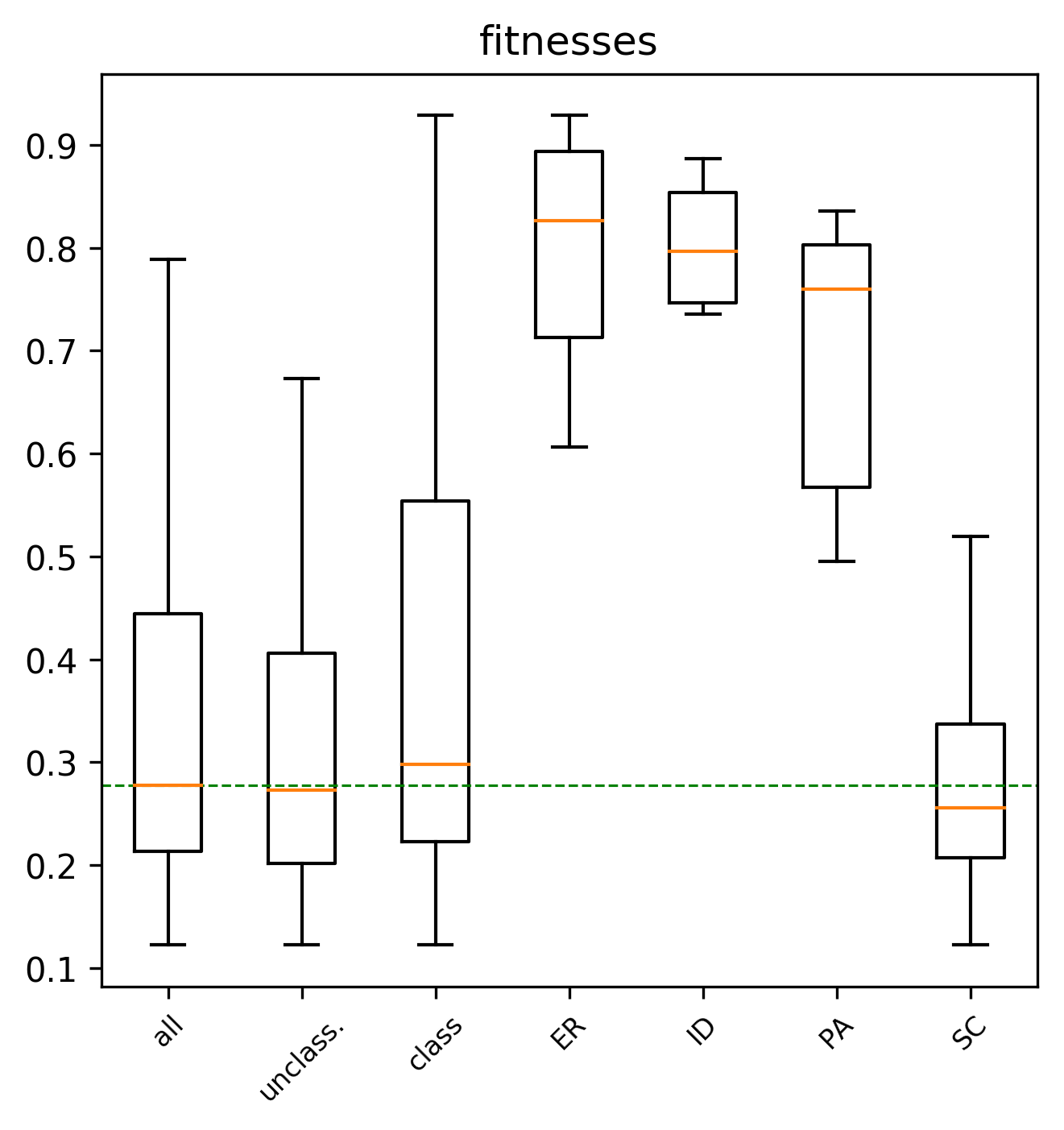}
\caption{Boxplots of best fitnesses achieved (lower is better) for the underlying networks of the various families, as well as all, unclassified and classified. Horizontal dashed line indicates overall median.}
\label{fig:fitnesses}
\end{figure}

In an attempt to test the second hypothesis, in figure~\ref{fig:fitnesses} we plot the best fitnesses achieved for the underlying networks, again per family, all generators, classified and unclassified. Here we find that generators of the \SC{} family attain slightly better fitness (both for the median and worst cases) than generators of unclassified networks. This lends some credence to the second hypothesis. Furthermore, we observed that for the entire \SC{} family, the generator with a simple pattern was only found once, and it always had the best fitness of the five runs. It seems thus likely that, given more evolutionary search runs per generator, at least part of the unclassified networks would fall into a family.



It is not possible to know if the families are exhaustive or the simplest that could be found. Investing more computational power on this problem could always yield simpler yet harder to find explanations, both for the classified and unclassified cases. It could also show unclassified networks to belong to a known family, or to a new family. As with many heuristic methods, the best we can do is trust some stability criteria (\hbox{e.g.}, stop at a certain number of runs without anything new being found).

\section{Final Remarks}
\label{sec:final}

We believe that several interesting explorations can stem from the symbolic regression of network generators. After the research work presented in this chapter, we are left encouraged by the potential of a genotype-based approach in describing families of generators. To move to a larger scale, it is necessary to go further in the methods to identify similar generators at the semantic, \hbox{i.e.} mathematical level. This is a hard but exciting computer science problem.

It would also be interesting to map the space of possible generators, by searching not for generators that target specific networks, but instead that attempt to generate networks as divergent as possible from those already known. Combining this exploration with family identification could lead to insights related to the families of generators found in different scientific fields and types of phenomena, as well as to families that do not correspond to networks found in any empirical data. This could reveal potentially interesting network designs, as was the case with the evolved radio antennas.

Our current method assumes homogeneous behavior across the network. Hybrid methods combining community detection with symbolic regression could lead, in certain cases, to more powerful explanations with different generator expressions per sub-network.

Another important challenge is that of targeting dynamic networks. This will require a fitness function that takes into account different stages of growth of a target network, and that leads to generators that can be validated to not only produce a plausible state of the network at a certain stage, but a plausible growth process overall.

\section*{Acknowledgements}
We are grateful to the members of our ``Algopol'' project team (ANR-12-CORD-0018) who organized most of the Facebook survey, including Ir\`ene Bastard, Dominique Cardon, Rapha\"el Charbey, Guilhem Fouetillou, Christo\-phe Prieur and St\'ephane Raux.
We further acknowledge interesting discussions with Jean-Philippe Cointet and David Fourquet regarding generative families, as well as the constructive feedback of our anonymous reviewers. This paper has been partially supported by the ``Algodiv'' grant (ANR-15-CE38-0001) funded by the ANR (French National Agency of Research).

\bibliographystyle{plainnat}
\bibliography{main}

\end{document}